\documentclass[a4paper,11pt]{article}
\usepackage{jheppub}
\usepackage{graphicx}  % needed for figures
\usepackage{caption}
\usepackage{subcaption}
\usepackage{amsmath, amssymb}   % for math
\usepackage{dsfont}
\usepackage{amsfonts}
\usepackage{xcolor}
\usepackage{hyperref}
\usepackage[applemac]{inputenc}
\usepackage{rotating}

\newcommand{\be}{\begin{equation}}
\newcommand{\ee}{\end{equation}}
\newcommand{\ba}{\begin{eqnarray}}
\newcommand{\ea}{\end{eqnarray}}
\newcommand{\nn}{\nonumber}

\newcommand{\lp}{\left(}
\newcommand{\rp}{\right)}
\newcommand{\ls}{\left[}
\newcommand{\rs}{\right]}

\newcommand{\N}{\mathcal{N}}
\newcommand{\Z}{\mathcal{Z}}
\renewcommand{\O}{\mathcal{O}}
\newcommand{\Tr}{{\rm Tr}}
\newcommand{\ch}{{\rm ch}}
\newcommand{\K}{K{\"a}hler}

\renewcommand{\Im}{{\rm Im}}
\newcommand{\rmi}{{\rm i}}
\newcommand{\CP}{\mathbb{CP}}

\definecolor{mygray}{rgb}{1, .5, .25}
\title{Calabi-Yau manifolds and sporadic groups} %to be improved

\author[]{Andreas  Banlaki,}
\author[]{Abhishek Chowdhury,}
\author[]{Abhiram Kidambi,}
\author[]{Maria Schimpf,}
\author[]{Harald Skarke,}
\author[]{and Timm Wrase}
\affiliation[]{Institute for Theoretical Physics, \\ TU Wien, \\Wiedner Hauptstr. 8-10, \\A-1040 Vienna, \\Austria\\}
\emailAdd{\\banlaki@hep.itp.tuwien.ac.at} % Is there a better email address?
\emailAdd{\\abhishek.chowdhury@tuwien.ac.at}
\emailAdd{\\abhiram.kidambi@tuwien.ac.at}
\emailAdd{\\mschimpf@hep.itp.tuwien.ac.at}
\emailAdd{\\skarke@hep.itp.tuwien.ac.at}
\emailAdd{\\timm.wrase@tuwien.ac.at}

\abstract{A few years ago a connection between the elliptic genus of the K3 manifold and the largest Mathieu group M$_{24}$ was proposed. We study the elliptic genera for Calabi-Yau manifolds of larger dimensions and discuss potential connections between the expansion coefficients of these elliptic genera and sporadic groups. While the Calabi-Yau 3-fold case is rather uninteresting, the elliptic genera of certain Calabi-Yau $d$-folds for $d>3$ have expansions that could potentially arise from underlying sporadic symmetry groups. We explore such potential connections by calculating twined elliptic genera for a large number of Calabi-Yau 5-folds that are hypersurfaces in weighted projected spaces, for a toroidal orbifold and two Gepner models.\vspace{3cm}
}

\notoc

\begin{document}
%\makeatletter
%\let\old@fpheader\@fpheader
%\renewcommand{\@fpheader}{%\hfill TUW-17-XX
%}
%\makeatother

\maketitle
	
\section{\label{sec:intro} Introduction}
In 2010 Eguchi, Ooguri and Tachikawa (EOT) \cite{EOT} discovered a new moonshine phenomenon \cite{MirandaTwining, Gaberdiel:2010ch, Gaberdiel:2010ca, EguchiTwining, GannonModule} which connects the elliptic genus of K3 to the largest Mathieu group M$_{24}$.\footnote{Relatedly, M$_{24}$ has at the same time been connected to the counting functions of half BPS states in four dimensional $\N=4$ theories \cite{Govindarajan:2009qt, MirandaTwining, Govindarajan:2010fu}.} Since the K3 surface plays a central role in mathematics and physics, this new observation has led to a flurry of papers (see \cite{Duncan:2014vfa, Kachru:2016nty} for recent review articles). Until today, the original EOT observation has not found a satisfying explanation but lots of progress has been made throughout the years and several new moonshine phenomena have been discovered and others have been better understood.

The K3 surface is the lowest dimensional non-trivial compact Calabi-Yau manifold. Its geometric symmetries at different points in moduli space have been classified by Mukai and Kondo \cite{Mukai, Kondo} who found that the symplectic automorphisms of any K3 manifold form a subgroup of the Mathieu group M$_{23}$. This means that no particular K3 surface has M$_{24}$ as symmetry group. However, the elliptic genus is an index and therefore it does not change if one moves around in moduli space. This has led to the intriguing idea that in order to get M$_{24}$ one could combine symmetries at different points in K3 moduli space. Using this so called `symmetry surfing' it has already been possible to substantially enlarge the symmetry group but it has not yet been possible to find the full M$_{24}$ group \cite{Taormina:2011rr, Taormina:2013jza, Taormina:2013mda, Gaberdiel:2016iyz}.

Following the discovery of Mathieu moonshine Gaberdiel, Hohenegger and Volpato \cite{Gaberdiel} showed that the supersymmetry preserving automorphisms of any nonlinear sigma model with K3 target generate a subgroup of the Conway group Co$_1$. This subgroup, however, is never M$_{24}$. The authors of \cite{Gaberdiel} also explicitly showed that the symmetry group at certain points in K3 moduli space, for example orbifold points, does not fit into M$_{24}$. Since the elliptic genus only counts BPS states of the theory, one might have hoped that the BPS states have M$_{24}$ as symmetry group, but the full spectrum has a different symmetry group. By calculating explicit twined elliptic genera \cite{Cheng:2016org,Paquette:2017gmb} one finds that there are more twined elliptic genera for K3 spaces than conjugacy classes of M$_{24}$. The symmetry groups of the K3 manifold at different points in moduli space have found a beautiful string theory explanation in \cite{Kachru:2016ttg}, where the authors study type IIA string theory on K3$\times T^3$. The resulting three dimensional theory does not only have M$_{24}$ symmetry at one point in moduli space but it actually has all umbral symmetry groups \cite{umbralone,umbraltwo} as symmetry groups at certain points in its moduli space. These symmetries are partially broken upon taking the decompactification limit for the $T^3$. Thus the elliptic genus of K3 has provided us with a fascinating but still not understood connection between the simplest non-trivial Calabi-Yau manifold and the largest
Mathieu group M$_{24}$.

The elliptic genus is an index that can be computed for any SCFT with $\N=(2,2)$ or more supersymmetry. In particular, one can calculate it for any conformal field theory with Calabi-Yau target space. The elliptic genus for a Calabi-Yau $d$-fold, where $d$ denotes the complex dimension of the Calabi-Yau manifold, is a Jacobi form with weight 0 and index $\frac{d}{2}$ \cite{Kawai:1993jk}. The space of such Jacobi forms, which we denote by $J_{0,\frac{d}{2}}$, is generated by very few basis elements for small $d$. Therefore the elliptic genus can carry only very little information about the actual Calabi-Yau $d$-fold. This does not mean that the elliptic genus is uninteresting, as is shown by the case of K3 where the space of weight zero, index 1 Jacobi forms is generated by a single function only. It does imply, however, that for larger $d$ many different Calabi-Yau $d$-folds will give rise to the same elliptic genus since the number of Calabi-Yau manifolds grows much faster with $d$ than the number of basis elements of $J_{0,\frac{d}{2}}$. Thus, if one finds interesting expansion coefficients in higher dimensional manifolds, then one faces the question of what this actually means. If the expansion coefficients are given in terms of irreducible representations of a particular sporadic group, does this imply that all manifolds with such elliptic genus are connected to the particular sporadic group, or only a few or none? How would such a connection manifest itself, as a geometric symmetry of the manifold or as a symmetry of the non-linear sigma model with the manifold as target space or via something else like symmetry surfing?

In this paper we try to identify interesting elliptic genera for Calabi-Yau $d$-folds with $d=5$ and larger (see \cite{CY4} for the case $d=4$).  The case of $d=5$ seems particularly promising: when expanding the elliptic genus in terms of $\N=2$ characters one finds essentially the same expansion coefficients as in Mathieu moonshine. Since the connection between Mathieu moonshine and the K3 surface is still somewhat mysterious, we may gain a clearer picture by studying CY$_5$ manifolds in order to understand whether they could be involved in Mathieu moonshine as well. By calculating twined elliptic genera for more than ten thousand explicit examples, we investigate the possibility that these Calabi-Yau 5-folds are connected to Mathieu or Enriques moonshine.

For $d>5$ potential connections to sporadic groups grow dramatically and the corresponding CY $d$-folds have not been constructed in any systematic way, so we refrain from a detailed investigation. However, we point out that a particular extremal Jacobi form that plays a role in umbral moonshine \cite{umbralone} arises as elliptic genus of the product of two CY 3-folds. We again study this further by looking at twined functions.

The paper is organized as follows: In section \ref{sec:Jacobi} we review some properties of Jacobi forms. Then, in section \ref{sec:elliptic}, we discuss the elliptic genus of Calabi-Yau manifolds with different dimensions as well as its expansion in $\N=2$ characters. In section \ref{sec:examples} we calculate twined elliptic genera for many different Calabi-Yau manifolds. Then we study a simple toroidal orbifold and two Gepner models in section \ref{sec:Gep}. We summarize our findings in section~\ref{sec:conclusion}. Our notation and conventions are given in appendix \ref{app:conventions}. Appendix \ref{app:characters} gives the $\N=2$ and $\N=4$ superconformal characters. Finally, appendix \ref{app:CharTab} contains the character tables for M$_{12}$ and M$_{24}$.

\section{\label{sec:Jacobi} Jacobi forms}
The elliptic genus of a CY $d$-fold, $Z_{CY_d}(\tau,z)$ with $\tau, % \in \mathbb{H}$ and $
z \in \mathbb{C}$ and $\Im(\tau) >0$, is a weight 0 and index $d/2$ weak Jacobi form \cite{Kawai:1993jk}. A Jacobi form $\phi_{k,m}(\tau,z)$ of weight $k$ and index $m$ satisfies %the following equations
\ba
\phi_{k,m} \lp \frac{a \tau+b}{c \tau+d},\frac{z}{c \tau+d}\rp &=& (c\tau+d)^k e^{\frac{2\pi \rmi m c z^2}{c\tau+d}} \phi_{k,m}(\tau,z)\,, \qquad\qquad \quad\lp \begin{array}{cc} a&b\\ c&d \end{array} \rp \in SL(2,\mathbb{Z})\,,\cr
\phi_{k,m}(\tau, z+\lambda \tau+\mu)&=&(-1)^{2 m(\lambda +\mu)} e^{-2\pi \rmi m (\lambda^2 \tau+2\lambda z)} \phi_{k,m}(\tau,z)\,, \qquad \lambda,\mu \in \mathbb{Z} \,.
\ea
The above transformations for $a=d=1$, $c=0$ and $\lambda=0$ allow us to do the following Fourier expansion
\be
\phi_{k,m}(\tau,z) = \sum_{n,r} c(n,r) q^n y^r\,,
\ee
with $q=e^{2 \pi \rmi \tau}$ and $y=e^{2 \pi \rmi z}$. A holomorphic Jacobi form satisfies $c(n,r)=0$ whenever $4nm-r^2<0$. What is usually called a weak Jacobi form satisfies the weaker condition $c(n,r)=0$ whenever $n<0$. We will be dealing with weak Jacobi forms in this paper but for simplicity we refer to them as Jacobi forms. 

The space of weak Jacobi forms \cite{bookZagier} of even weight $k$ and integer index $m$ is generated by the four functions $E_{4}(\tau), E_6(\tau), \phi_{-2,1}(\tau,z), \phi_{0,1}(\tau,z)$ that we define in appendix \ref{app:conventions}.\footnote{Forms of odd weight have one extra generator $\phi_{-1,2} = \rmi \frac{\theta_1(\tau,2z)}{\eta(\tau)^3}$ that satisfies $432 \phi_{-1,2}^2 = - \phi_{-2,1} \lp \phi_{0,1}^3 -3E_4 \phi_{-2,1}^2 \phi_{0,1}-2 E_6 \phi_{-2,1}^3\rp$, which is not relevant for even weight.} The Eisenstein series $E_4$ and $E_6$ have weight 4 and 6 respectively and vanishing index. %The funtions $\phi_{k,m}$ have weight $k$ and index $m$.
It then follows from simple combinatorics that the space $J_{0,m}$ of Jacobi forms of weight 0 and index $m$, is generated by $m$ functions for $m=1,2,3,4,5$. In particular, we have
\ba
J_{0,1}&=& \langle  \phi_{0,1} \rangle\,,\cr
J_{0,2}&=& \langle  \phi_{0,1}^2, E_4\, \phi_{-2,1}^2 \rangle\,,\cr
J_{0,3}&=& \langle  \phi_{0,1}^3, E_4\, \phi_{-2,1}^2\, \phi_{0,1}, E_6\, \phi_{-2,1}^3 \rangle\,,\cr
J_{0,4}&=& \langle  \phi_{0,1}^4, E_4\, \phi_{-2,1}^2\, \phi_{0,1}^2, E_6\, \phi_{-2,1}^3\, \phi_{0,1},E_4^2\, \phi_{-2,1}^4 \rangle\,,\cr
J_{0,5}&=& \langle  \phi_{0,1}^5, E_4\, \phi_{-2,1}^2\, \phi_{0,1}^3, E_6\, \phi_{-2,1}^3\, \phi_{0,1}^2,E_4^2\, \phi_{-2,1}^4\, \phi_{0,1}, E_4 \, E_6\, \phi_{-2,1}^5 \rangle\,.\label{eq:JacBasis}
\ea
The functions above appear in the elliptic genus of Calabi-Yau $d=2,4,6,8,10$ manifolds and their coefficients can be fixed in terms of a few topological numbers of the CY $d$-fold.

For weight zero Jacobi forms with half integer index things are equally simple. In general, for any even weight the spaces of Jacobi forms with integer and half integer indices are related as follows (see for example Lemma 1.4 in \cite{Gritsenko:1999fk})
\be
J_{2k,m+\frac12} = \phi_{0,\frac32} J_{2k,m-1}\,,\qquad m \in \mathbb{Z}\,.
\ee
%This means i
In particular $\phi_{0,\frac32}$ and $\phi_{0,\frac32} \phi_{0,1}$ are, up to rescaling, the unique Jacobi forms of weight 0 and index $\frac32$ and $\frac 52$, respectively. More generally, the space $J_{0,m+\frac32}$ is spanned by $m$ functions for $m=1,2,3,4,5$ and these functions are the ones given in equation \eqref{eq:JacBasis} multiplied by $\phi_{0,\frac32}$.\footnote{Note that we have the identity $432\phi_{0,\frac32}^2= \phi_{0,1}^3-3 E_4\, \phi_{-2,1}^2\, \phi_{0,1}-2 E_6\, \phi_{-2,1}^3 $.} So in summary we see that the space of Jacobi forms $J_{0,\frac{d}{2}}$ is generated by very few functions for small $d$.

\section{\label{sec:elliptic} The elliptic genera for Calabi-Yau manifolds}
Here we gather some results on the elliptic genera of Calabi-Yau manifolds in different dimensions (see for example \cite{Gritsenko:1999fk, Eguchi:2010xk} for a related discussion). %The for a
A superconformal $\N=(2,2)$ theory with CY$_d$ target space has central charges of $(c,\bar c) = (3d,3d)$ and therefore its elliptic genus is given by \cite{Edold}
\be\label{eq:ellgenus}
\Z_{CY_d}(\tau,z) = \Tr_{RR} \lp (-1)^{F_L} y^{J_0} q^{L_0-\frac{d}{8}} (-1)^{F_R} \bar{q}^{\bar L_0-\frac{d}{8}}\rp
\ee
in the standard notation $q=e^{2 \pi \rmi \tau}$ and $y=e^{2 \pi \rmi z}$.
%we used the fact that the .
The above index receives only contributions from the right moving ground state and therefore does not depend on $\bar q$.

The elliptic genus has the property
\be\label{eq:ellprop}
\Z_{CY_d}(\tau,z) = \sum_{p=0}^d (-1)^p \chi_p(CY_d) y^{\frac{d}{2}-p} + \O(q)\,,
\ee
where $\chi_p(CY_d) = \sum_{r=0}^d (-1)^r h^{p,r}$. The higher order terms in $q$ vanish for $z=0$ ($y=1$) since the elliptic genus reduces in this case to the Witten index (cf. equation \eqref{eq:ellgenus}). This implies that 
\be\label{eq:elleuler}
\Z_{CY_d}(\tau,0) = \chi_{CY_d} = \sum_{p=0}^d (-1)^p  \chi_p(CY_d)
\ee
is exactly the Euler number of the Calabi-Yau. These properties of the elliptic genus allow us to fix the prefactors of the generators of $J_{0,\frac{d}{2}}$ in terms of the Euler number of the Calabi-Yau and maybe one or two $\chi_p$ for small $d$.

\subsection{Calabi-Yau 1-folds}
For the standard torus $T^2$ the elliptic genus vanishes, $\Z_{T^2}(\tau,z) = 0$. The same holds true for any even dimensional torus $\Z_{T^{2n}}(\tau,z) = 0$, $\forall n \in \mathbb{N}$. This is due to the fermionic zero modes in the right moving Ramond sector: $\Tr \lp (-1)^{F_R} \bar q^{\bar{L}_0-\frac{\bar c}{24}} \rp \propto \theta_2(\bar q,-1)^n = 0$.

\subsection{Calabi-Yau 2-folds}
The elliptic genus for a K3 surface is the Jacobi form that appears in Mathieu moonshine. %and we have the following elliptic genus
Its expansion in terms of $\N=4$ characters \cite{EOTY} is\footnote{See appendix \ref{app:conventions} for standard definitions and our conventions and appendix \ref{app:characters} for the explicit formulas for the $\N=2$ and $\N=4$ characters. The subscripts on the characters $\ch_{d,h-c/24,\ell}$ are the complex dimension $d$ of the Calabi-Yau that determines the central charge to be $c=3d$, the eigenvalue $h-c/24$ of $L_0-c/24$ and the eigenvalue $\ell$ of $J_0$.}
\be
\Z_{K3}(\tau,z) = 2 \phi_{0,1}(\tau,z) = 20\,\ch_{2,0,0}^{\N=4}(\tau,z) -2\, \ch_{2,0,\frac12}^{\N=4}(\tau,z) + \sum_{n=1}^\infty A_n\ch_{2,n,\frac12}^{\N=4}(\tau,z)\,,
\ee
where all coefficients can be expressed in a simple way via irreducible representations of M$_{24}$. In particular, we have
\ba
20&=&23-3 \cdot 1\,,\cr
-2&=&-2 \cdot 1\,,\cr
A_1 &=&45+\underline{45}\,,\cr
A_2 &=&231 + \underline{231}\,,\cr
A_3 &=&770+\underline{770}\,,\cr
&\ldots&\,,\label{eq:M24decom}
\ea
where the numbers on the right hand side are dimensions of irreducible representations of M$_{24}$ (see appendix \ref{app:CharTab}) and we used the underline to distinguish two different irreducible representations of the same dimension. Note that the elliptic genus counts states with sign and all coefficients could have in principle been positive or negative but they turn out to all be positive except one.

Above we have expanded the elliptic genus in terms of $\N=4$ characters following the original work \cite{EOT}. This can be done since K3 is a hyper {\K} manifold and the superconformal worldsheet theory has $\N=(4,4)$ supersymmetry. However, we can also expand the elliptic genus of K3 in terms of the $\N=2$ characters given in \cite{Eguchi:2010xk} \footnote{For the particular case of $c=6$, i.e. for $d=2$, the $\N=2$ superconformal algebra extended by spectral flow generators is the same as the $\N=4$ superconformal algebra. This means that in this case the $\N=2$ and the $\N=4$ algebras are the same (up to an overall sign in our conventions).}
\be
\Z_{K3}(\tau,z) = 2 \phi_{0,1}(\tau,z) = -20\,\ch_{2,0,0}^{\N=2}(\tau,z) +2\, \ch_{2,0,1}^{\N=2}(\tau,z) - \sum_{n=1}^\infty A_n\ch_{2,n,1}^{\N=2}(\tau,z)\,.
\ee
So up to an overall sign we see the same expansion coefficients as before. This overall sign is a convention (see the sentence after equation (2.5) in \cite{Eguchi:2010xk}).

There is also a moonshine phenomenon that connects Enriques surfaces to M$_{12}$ \cite{Eguchi:2013es}. The elliptic genus for an Enriques surface is $\Z_{\rm Enr}(\tau,z) =\frac12 \Z_{K3}(\tau,z)$, since $\chi_{\rm Enr} =12= \Z_{\rm Enr}(\tau,0)=\phi_{0,1}(\tau,0)$ (cf. equations \eqref{eq:elleuler} and \eqref{eq:phi01}). This leads to the following expansion in terms of $\N=4$ characters
\be
\Z_{\rm Enr}(\tau,z) = \phi_{0,1}(\tau,z) = 10\,\ch_{2,0,0}^{\N=4}(\tau,z) -1\, \ch_{2,0,\frac12}^{\N=4}(\tau,z) + \sum_{n=1}^\infty \frac{A_n}{2} \ch_{2,n,\frac12}^{\N=4}(\tau,z)\,.
\ee
Note that all the $A_n$ are even so that also in this case the expansion coefficients are always integers and one can again decompose the coefficients into irreducible representations of -- this time -- M$_{12}$:
\ba
10 &=&11-1\,,\cr
-1&=&-1\,,\cr
A_1 &=& 45\,,\cr
A_2 &=& 55 + 176\,,\cr
A_3 &=& 66+2\cdot 120+2\cdot 144+176\,,\cr
&\ldots&\,.\label{eq:M12decom}
\ea
We can of course again expand the elliptic genus $\Z_{\rm Enr}(\tau,z)$ in $\N=2$ characters, which leads to the same expansion coefficients but with an overall minus sign.

\subsection{Calabi-Yau 3-folds}
Calabi-Yau 3-folds have a rather uninteresting expansion of the elliptic genus in terms of $\N=2$ characters
\be
\Z_{CY_3}(\tau,z) = \frac{\chi_{CY_3}}{2} \ \phi_{0,\frac32} = \frac{\chi_{CY_3}}{2} \lp \ch_{3,0,\frac12}^{\N=2}(\tau,z)+ \ch_{3,0,-\frac12}^{\N=2}(\tau,z) \rp\,,
\ee
where the overall normalization is fixed via equations \eqref{eq:elleuler} and \eqref{eq:phi032}. Only two expansion coefficients are different from zero and they are both equal. So %the expansion coefficients and therefore
the elliptic genus carries little information.

While we study CY 3-folds only very briefly in this paper, we would like to stress that the simplicity of the elliptic genus does not mean that Calabi-Yau 3-folds cannot be interesting or have no connection to moonshine phenomena. For example, it was shown in \cite{Cheng:2013kpa} that the Gromov-Witten invariants of certain CY$_3$ manifolds are connected to Mathieu moonshine. This was further explored and studied in \cite{Wrase:2014fja, Paquette:2014rma, Datta:2015hza, Chattopadhyaya:2016xpa, Chattopadhyaya:2017ews}.

\subsection{Calabi-Yau 4-folds}\label{ssec:CY4}
For Calabi-Yau 4-folds the corresponding space $J_{0,2}$ of Jacobi forms is generated by two basis elements and one can fix the coefficients in terms of the Euler number $\chi_{CY_4}$ and $\chi_0=\sum_r (-1)^r h^{0,r}$ using equations \eqref{eq:ellprop} and \eqref{eq:elleuler} as well as the explicit forms of the basis elements given in appendix \ref{app:conventions}. One then finds
\be
\Z_{CY_4}(\tau,z) = \frac{\chi_{CY_4}}{144} \lp \phi_{0,1}^2-E_4\, \phi_{-2,1}^2 \rp +\chi_0 E_4\, \phi_{-2,1}^2\,.
\ee
Note that $\chi_0=h^{0,0}+h^{0,4}=2$ for genuine CY $4$-folds.

There are a variety of interesting connections to sporadic groups that have already appeared in the literature. The first and somewhat trivial case is the product of two K3 manifolds whose elliptic genus is given by $\Z_{K3\times K3}(\tau,z) = 4 \phi_{0,1}^2$. This function exhibits an M$_{24} \times$M$_{24}$ symmetry and one could wonder whether such a symmetry (or an M$_{12}\times$M$_{24}$ symmetry for $2 \phi_{0,1}^2$) is realized by certain genuine Calabi-Yau four manifolds. The Jacobi form $\frac{1}{24}\lp \phi_{0,1}^2-E_4\, \phi_{-2,1}^2 \rp$ has appeared in umbral moonshine \cite{umbralone} where it was shown to exhibit $2.M_{12}$ moonshine when expanded in $\N=4$ characters, and in \cite{Eguchi:2012ye} a $L_2(11)$ moonshine was established upon expanding this function in $\N=2$ characters. Lastly in \cite{Cheng:2014owa} it was shown that the function $\frac16(\phi_{0,1}^2+ 5 E_4\, \phi_{-2,1}^2)$ exhibits M$_{22}$ moonshine when expanded in $\N=4$ characters and M$_{23}$ moonshine when expanded in $\N=2$ characters.\footnote{This functions also exhibits M$_{24}$ moonshine when expanded in extended $\N=1$ characters \cite{Cheng:2015fha}. There are also other groups that can arise instead of the Mathieu groups but the later are somewhat special \cite{Cheng:2014owa, Cheng:2015fha}. When evaluated at torsion points $y=-1$ and $y=\pm \sqrt{q}$, the same function gives rise to Conway moonshine \cite{Frenkel1985, Duncan}.} All of these and further potential connections to sporadic groups will be discussed in \cite{CY4} where also twinings will be studied in great detail.

\subsection{Calabi-Yau 5-folds}
As discussed above, the elliptic genus for a Calabi-Yau manifold with odd complex dimensions $d$ can be expressed in terms of the same functions as those occurring in the expression for the elliptic genus of a Calabi-Yau manifold with complex dimension $d-3$, since Jacobi forms of half-integral index can be written as the product of $\phi_{0,\frac32}$ times an integral index Jacobi form $J_{0,\frac{d}{2}}=\phi_{0,\frac32} J_{0,\frac{d-3}{2}}$ \cite{Gritsenko:1999fk}. This means that the elliptic genus for CY 5-folds is proportional to $\phi_{0,\frac32} \phi_{0,1}$ and we can fix the prefactor in terms of the Euler number $\chi_{CY_5}$. We also recall that the $\N=4$ characters for central charge $c=3d$, multiplied by $\phi_{0,\frac32}$, can be expressed in terms of $\N=2$ characters for central charge $c=3(d+3)$  (see appendix \ref{app:characters}). In particular, we find the following relations
\ba
\phi_{0,\frac32} \ch^{\N=4}_{2,0,0} &=& -\ch^{\N=2}_{5,0,\frac12}-\ch^{\N=2}_{5,0,-\frac12}\,,\cr
\phi_{0,\frac32} \ch^{\N=4}_{2,0,\frac12} &=& -\ch^{\N=2}_{5,0,\frac32}-\ch^{\N=2}_{5,0,-\frac32}+\ch^{\N=2}_{5,0,\frac12}+\ch^{\N=2}_{5,0,-\frac12}\,,\cr
\phi_{0,\frac32} \ch^{\N=4}_{2,n,\frac12} &=& -\ch^{\N=2}_{5,n,\frac32}-\ch^{\N=2}_{5,n,-\frac32}\,,\quad \forall n=1,2,\ldots\,.\label{eq:N4toN2}
\ea
This means that CY 5-folds have the following expansion of the elliptic genus in terms of $\N=2$ characters:
\ba\label{eq:CY5ell}
\Z_{CY_5}(\tau,z) &=& \frac{\chi_{CY_5}}{24} \phi_{0,\frac32} \phi_{0,1}\cr
&=& -\frac{\chi_{CY_5}}{48} \Bigg[ 22\lp \ch_{5,0,\frac12}^{\N=2}(\tau,z)+\ch_{5,0,-\frac12}^{\N=2}(\tau,z) \rp -2\lp\ch_{5,0,\frac32}^{\N=2}(\tau,z)+\ch_{5,0,-\frac32}^{\N=2}(\tau,z)\rp \cr
&&\qquad\qquad+\sum_{n=1}^\infty A_n \lp \ch_{5,n,\frac32}^{\N=2}(\tau,z)+\ch_{5,n,-\frac32}^{\N=2}(\tau,z) \rp\Bigg]\,.
\ea
In particular, for CY 5-folds with $\chi_{CY_5}=-48$ we find essentially the same expansion coefficients as in Mathieu moonshine, while for $\chi_{CY_5}=-24$ we find essentially the same coefficients as for Enriques moonshine. Since the overall sign in the definition of the $\N=2$ characters is a choice, we can conclude the same for CY 5-folds with $\chi_{CY_5}=48$ and $\chi_{CY_5}=24$. The decomposition of all the coefficients into the dimensions of irreducible representations is identical to the ones given above in equations \eqref{eq:M24decom} and \eqref{eq:M12decom} except for the first coefficient for which we have $22=23-1$ for M$_{24}$ and $11=11$ for M$_{12}$. This follows from the relations between $\N=4$ and $\N=2$ characters given above in equation \eqref{eq:N4toN2}.

A trivial class of examples where the elliptic genus will exhibit Mathieu or Enriques moonshine is given by manifolds that are products of K3 or Enriques surfaces with CY 3-folds. While these cases are rather trivial, we see from the above that any CY 5-fold could exhibit an interesting connection to a Mathieu group. Given our current still incomplete understanding of Mathieu moonshine we find it important to study this further by calculating twined elliptic genera for genuine CY 5-folds.

\subsection{Calabi-Yau 6-folds}
The elliptic genus for Calabi-Yau 6-folds is determined in terms of three topological numbers which we choose to be the Euler number $\chi_{CY_6}$ and $\chi_p = \sum_{r=0}^6 (-1)^r h^{p,r}$ for $p=0,1$. The elliptic genus is then given by
\ba
\Z_{CY_6}(\tau,z) &=& \frac{\chi_{CY_6}}{1728} \phi_{0,1}^3 -\frac{1}{576}(\chi_{CY_6} - 48 (\chi_1 + 6 \chi_0)) E_4\, \phi_{-2,1}^2\, \phi_{0,1}\cr
&&\qquad  \qquad\ -\frac{1}{864} (\chi_{CY_6} - 72 (\chi_1 - 6 \chi_0))  E_6\, \phi_{-2,1}^3\,.
\ea
We can again find many interesting cases by taking triple products of $K3$ or Enriques surfaces or by considering $K3\times CY_4$. These cases trivially exhibit a variety of potential connections to sporadic groups like L$_2(11)$, M$_{12}$, M$_{22}$, M$_{23}$ and M$_{24}$ (see subsection \ref{ssec:CY4}). Again it would be interesting to understand whether genuine Calabi-Yau 6-folds with the appropriate topological numbers also have such a connection to a sporadic symmetry group.

For $d=6$ there is also one genuinely new case: for $\chi_0=\chi_1=0$ we get
\ba\label{eq:umbral6}
\Z_{CY_6}(\tau,z) &=& \frac{\chi_{CY_6}}{1728} \lp \phi_{0,1}^3 -3 E_4\, \phi_{-2,1}^2\, \phi_{0,1}-2  E_6\, \phi_{-2,1}^3 \rp \cr
&=&\frac{\chi_{CY_6}}{4} \phi_{0,\frac32}^2\cr
&=&  \chi_{CY_6} \frac{\theta_2(\tau,z)^2}{\theta_2(\tau,0)^2}\frac{\theta_3(\tau,z)^2}{\theta_3(\tau,0)^2}\frac{\theta_4(\tau,z)^2}{\theta_4(\tau,0)^2}  \cr
&=& \frac{ \chi_{CY_6}}{8} \Big[ 4 \ch^{N=4}_{6,0,0} + \lp -2 \ch^{N=4}_{6,0,\frac12} + 14 \ch^{N=4}_{6,1,\frac12} + 42 \ch^{N=4}_{6,2 ,\frac12}+ 86 \ch^{N=4}_{6,3 ,\frac12} +\ldots \rp\cr
&&\qquad \qquad \qquad\,  - \lp 16 \ch^{N=4}_{6,1,1} + 48 \ch^{N=4}_{6,2 ,1}+ 112 \ch^{N=4}_{6,3 ,1} +\ldots \rp\cr
&&\qquad \qquad \qquad \,+ \lp 6 \ch^{N=4}_{6,1,\frac32} + 28 \ch^{N=4}_{6,2 ,\frac32}+ 56 \ch^{N=4}_{6,3 ,\frac32} +\ldots \rp\Big]\,.\quad
\ea
This Jacobi form for $\chi_{CY_6}=8$ and the corresponding expansion in terms of $\N=4$ characters appeared in \cite{umbralone} (see eqn. (2.49)), and it is related to $2.AGL_3(2)$ via the umbral moonshine conjecture (proven in \cite{umbralproof}). This is similar to the case of CY $4$-folds that seem connected to umbral moonshine for $\chi_0(CY_4)=0$. Here, however, we can easily find spaces that have the above elliptic genus. Since the product of two CY 3-folds gives a complex six dimensional manifold with $\chi_0=\chi_1=0$, it results in the elliptic genus above with $\chi_{CY_6} = \chi_{CY_3^{(1)}} \cdot \chi_{CY_3^{(2)}}$. For umbral moonshine we require an elliptic genus as given by equation \eqref{eq:umbral6} for the case of $\chi_{CY_6}=8$. So the product of a CY 3-fold with Euler number $\pm 2$ with another CY 3-fold with Euler number $\pm 4$ gives the desired answer.

\subsection{Calabi-Yau manifolds of dimension $d>6$}
It should be clear now that the potential connections to sporadic groups grow rapidly for larger $d$. There is no simple way of checking these, however, since CY $d$-folds for larger $d$ have not been constructed systematically and it would also be much more time consuming to study them with the methods employed in this paper. Nevertheless we can check under what conditions the extremal Jacobi forms that appear in umbral moonshine \cite{umbralone, umbraltwo} can arise as elliptic genera of Calabi-Yau manifolds or products thereof.

Recall that the elliptic genus of K3, i.e. a Jacobi form of weight 0 and index 1, provides the first example of umbral moonshine. As we mentioned above, the index 2 Jacobi form that arises in umbral moonshine would correspond to the elliptic genus of a CY $4$-fold with $\chi_0(CY_4)=0$. However, any genuine CY $d$-fold, by which we mean a manifold whose holonomy group is $SU(d)$ and not a subgroup thereof,  satisfies
\be
\chi_0(CY_d)=\left\{ \begin{array}{c} 0 \quad \text{if $d$ is odd,}\\ 2 \quad \text{if $d$ is even.} \end{array}\right.
\ee
So we cannot get the extremal Jacobi form of weight 0 and index 2 form from a genuine CY $4$-fold, nor from $K3 \times K3$ since $\chi_0(K3\times K3)=\chi_0(K3)^2=4$ ($T^4\times K3$ has vanishing $\chi_0$ since $\chi_0(T^4)\cdot \chi_0(K3)=0\cdot2=0$ but also vanishing elliptic genus). Above we have seen, however, that six complex dimensional Calabi-Yau spaces with $\chi_0=\chi_1=0$ have as elliptic genus the extremal Jacobi form of weight 0 and index 3 that appears in umbral moonshine. In particular the product of any two CY 3-folds has $\chi_0=\chi_1=0$.\footnote{Recall the useful formula $\chi_i(CY\times CY') = \sum_{j=0}^i \chi_j(CY)\cdot\chi_{i-j}(CY')$.} The other interesting Jacobi forms in \cite{umbralone} have weight zero and index 4, 6 and 12. Let us look first at the case of $d=8$ that leads to a Jacobi form with index 4. The relevant form in umbral moonshine arises for $\chi_0=\chi_1=\chi_2=0$. We can again ask whether we can get this from a product of CY manifolds. Unfortunately the answer is no. For the case of CY$_5\times$CY$_3$ we find $\chi_0=\chi_1=0$ but $\chi_2(CY_3\times CY_5) \propto \chi(CY_3)\cdot \chi(CY_5)$. So only if either $\chi(CY_3)=0$ or $\chi(CY_5)=0$, do we satisfy the required property. However, in that case we have $\Z_{CY_3\times CY_5}(\tau,z)\propto  \chi(CY_3)\cdot \chi(CY_5)=0$. Similarly for the case $K3\times CY_3 \times CY'_3$ we find $\chi_0=\chi_1=0$ but $\chi_2(K3\times CY_3\times CY'_3) \propto \chi(CY_3)\cdot \chi(CY'_3)$ and $\Z_{K3\times CY_3\times CY'_3}(\tau,z)\propto  \chi(CY_3)\cdot \chi(CY'_3)$ and there is again no non-trivial solution.

The same holds true for $d=12$ in which case the umbral moonshine Jacobi form arises as elliptic genus for spaces with $\chi_0=\chi_1=\chi_2=\chi_3=\chi_4=\chi_6=0$ which is very restrictive and does not seem to be realizable by taking products of Calabi-Yau manifolds. The most promising case, which is the product of four CY 3-folds, has $\chi_0=\chi_1=\chi_2=\chi_3=0$ but $\chi_4\neq 0 \neq \chi_6$ unless the Euler number of one of CY 3-folds vanishes in which case the entire elliptic genus is zero as well.

The original umbral moonshine was extended in \cite{umbraltwo} and there are many further Jacobi forms that could in principle arise from products of CY manifolds. However, again this does not seem possible for the cases we checked. For example, there is an extremal Jacobi form that could arise from the elliptic genus of a CY 10-fold with $\chi_0=\chi_1=\chi_2=\chi_3=0$, but again this cannot be realized by taking products of, for example, two K3 surfaces and two CY 3-folds. Similarly the index 9 extremal Jacobi form in umbral moonshine would correspond to a Calabi-Yau manifold with $\chi_p=0,~\forall p\leq 7$, but the product of six CY 3-folds has $\chi_p=0,~\forall p\leq 5$, and $\chi_6 \neq 0$, if we require the elliptic genus of this product to be non-vanishing. So it seems that only the extremal Jacobi forms of index 1 and 3 can arise as elliptic genera of Calabi-Yau manifolds or their products.

\section{\label{sec:examples} The twined elliptic genus for specific examples}
Above we have shown that the elliptic genus for Calabi-Yau manifolds with $d>3$ %is also interesting. It
often exhibits interesting expansions in terms of $\N=2$ and sometimes even $\N=4$ characters, where the expansion coefficients are related to irreducible representations of sporadic groups. An obvious question is then: What does this actually mean?

While we do not know the answer we can speculate that it is connected to understanding Mathieu moonshine. By analogy with K3 we expect that not all Calabi-Yau 5-folds with $\chi_{CY_5}=-48$ will have an M$_{24}$ symmetry at every point in their moduli space. Perhaps they do not even have such a symmetry at any point in their moduli space at all. However, the elliptic genus could see some combined symmetry group that arises, for example, from different points in moduli space via symmetry surfing, again in analogy with K3 \cite{Taormina:2013jza, Taormina:2013mda, Gaberdiel:2016iyz}. On the other hand it is not impossible that some CY 5-folds have a genuine M$_{24}$ symmetry at some point in the moduli space and that this could then explain the corresponding expansion of the elliptic genus. Something similar could happen for the other cases that involve Calabi-Yau 4-folds and 6-folds. It would be interesting to study this by classifying all the symmetry groups that arise for certain Calabi-Yau manifolds with $d=4,5$ or 6. However, this seems like a daunting task so we will here settle for the simpler task of explicitly calculating twined elliptic genera for many different Calabi-Yau manifolds that are hypersurfaces in weighted projective spaces.

\subsection{Calculating the twined elliptic genus}

We use the methods developed in \cite{Benini} (first applied to moonshine in \cite{Harrison:2013bya}) to calculate elliptic genera twined by a symmetry of the Calabi-Yau manifold. The Calabi-Yau manifolds we are interested in are hypersurfaces in weighted projective ambient spaces. A particular Calabi-Yau $d$-fold that is a hypersurface in the weighted projective space $\CP^{d+1}_{w_1\ldots w_{d+2}}$ is determined by a solution of $p(\Phi_1,\ldots,\Phi_{d+2})=0$, where the $\Phi_i$ denote the homogeneous coordinates of the weighted projective space and $p$ is a transverse polynomial of degree $m=\sum_i w_i$. We will now quickly review the results of \cite{Benini} on how to calculate the elliptic genus for such manifolds.

We consider a two-dimensional gauged linear sigma model with $\N=(2,2)$ supersymmetry. We have a $U(1)$ gauge field under which the chiral multiplets $\Phi_i$ have charge $w_i$. Additionally we have one extra chiral multiplet $X$ with $U(1)$ charge $-m$. The superpotential that is invariant under $U(1)$ gauge transformations is given by $W=X p(\Phi_1,\ldots,\Phi_{d+2})$. Then the F-term equation $\partial W/\partial X= p =0$ restricts us to the Calabi-Yau hypersurface above. We can assign $\mathcal{R}$-charge zero to the $\Phi_i$ and 2 to the chiral multiplet $X$ which ensures that the superpotential has always the correct $\mathcal{R}$-charge 2.

One can define a refined elliptic genus, depending on the extra chemical potential $x = e^{2 \pi \rmi u}$, as
\be\label{eq:refellgenus}
\Z_{\rm ref}(\tau,z, u) = \Tr_{RR} \lp (-1)^{F_L} y^{J_0} q^{L_0-\frac{d}{8}} x^Q (-1)^{F_R} \bar{q}^{\bar L_0-\frac{d}{8}}\rp\,.
\ee
This refined elliptic genus also keeps track of the $U(1)$ charges $Q$ of the states in the theory. Each chiral multiplet of $U(1)$ charge $Q$ and $\mathcal{R}$-charge $R$ gives a multiplicative contribution to this refined elliptic genus that is
\be
\Z^{\Phi}_{\rm ref}(\tau,z, u)  = \frac{\theta_1\lp\tau,\lp\tfrac{R}{2}-1\rp z+ Q u\rp}{\theta_1\lp\tau, \tfrac{R}{2} z+Q u\rp}\,.
\ee
Our abelian vector field leads to a ($u$ independent) factor
\be
\Z^{vec}_{\rm ref}(\tau,z)  = \frac{\rmi \eta(\tau)^3}{\theta_1(\tau, - z)}\,.
\ee
Combining these we find for our theory with
\begin{itemize}
\item $d+2$ chiral multiplets $\Phi_i$ with $U(1)$ charge $w_i$ and zero $\mathcal{R}$-charge, 
\item one chiral multiplet $X$ with $U(1)$ charge $-m$ and $\mathcal{R}$-charge 2
\item and one Abelian vector multiplet
\end{itemize}
the following refined elliptic genus
\be
\Z_{\rm ref }(\tau,z,u) = \frac{\rmi \eta(\tau)^3}{\theta_1(\tau,-z)} \frac{\theta_1\lp\tau, -m u\rp}{\theta_1\lp\tau, z-m u\rp}\prod_{i=1}^{d+2} \frac{\theta_1\lp \tau, -z+w_i u\rp}{\theta_1\lp \tau ,w_i u\rp} \,.
\ee
The standard elliptic genus is obtained by integrating over $u$. This integral localizes to a sum of contour integrals \cite{Benini} so that we have
\ba
\Z_{CY_d}(\tau,z) = \frac{\rmi \eta(\tau)^3}{\theta_1(\tau,-z)} \sum_{u_j\in \mathcal{M}_{sing}^-} \oint_{u=u_j} du \frac{\theta_1\lp\tau, -m u\rp}{\theta_1\lp\tau, z-m u\rp}\prod_{i=1}^{d+2} \frac{\theta_1\lp \tau, -z+w_i u\rp}{\theta_1\lp \tau ,w_i u\rp}\,,
\ea
where $\mathcal{M}_{sing}^-$ is the space of poles of the integrand where the chiral multiplets become massless. For a chiral multiplet with $U(1)$ charge $Q$ and $\mathcal{R}$-charge $R$ these singularities are located at
\ba
Q u+\frac{R}{2}z= 0\,, \quad \text{mod  } \mathbb{Z}+\tau \mathbb{Z}\,.
\ea
In the above formula one can restrict to the singularities for chiral multiplets with $Q<0$, hence the superscript $\mathcal{M}_{sing}^-$. In our particular setup only the chiral multiplet $X$ has negative $U(1)$ charge so the singularities are solutions to
\be
-m u +z =-k-\ell \tau\,, \quad k,\ell \in \mathbb{Z}\,.
\ee
The integrand above is periodic under the identification $u\sim u+1 \sim u+\tau$ and the solutions within one fundamental domain of $u$ are
\be
u  = (z+k+\ell \tau)/m\,, \quad 0\leq k,\ell<m\,.
\ee
So we can rewrite the above expression as
\ba
\Z_{CY_d}(\tau,z) = \frac{\rmi \eta(\tau)^3}{\theta_1(\tau,-z)}  \sum_{k,\ell=0}^{m-1}  \oint_{u=(k+\ell \tau+z)/m} du \frac{\theta_1\lp\tau, -m u\rp}{\theta_1\lp\tau, z-m u\rp}\prod_{i=1}^{d+2} \frac{\theta_1\lp \tau, -z+w_i u\rp}{\theta_1\lp \tau ,w_i u\rp}\,.\quad
\ea
The above can be further simplified by using properties of the $\theta$-function (see appendix B of \cite{Benini} for details). This leads to the following simple formula for the elliptic genus of a Calabi-Yau $d$-manifold that is a hypersurface in a weighted projective space and that can be described by a transverse polynomial:
\ba
\Z_{CY_d}(\tau,z) &=& \sum_{k,\ell=0}^{m-1} \frac{e^{-2\pi \rmi \ell z}}{m} \prod_{i=1}^{d+2} \frac{\theta_1\lp\tau , \frac{w_i}{m} (k+\ell \tau +z)-z\rp}{\theta_1\lp\tau , \frac{w_i}{m} (k+\ell \tau +z)\rp} \cr
&=& \sum_{k,\ell=0}^{m-1} \frac{y^{-\ell}}{m} \prod_{i=1}^{d+2} \frac{\theta_1\lp q, e^{\frac{2 \pi \rmi w_i k}{m}} q^{\frac{w_i \ell}{m}} y^{\frac{w_i}{m}-1}\rp}{\theta_1\lp q, e^{\frac{2 \pi \rmi w_i k}{m}} q^{\frac{w_i \ell}{m}} y^{\frac{w_i}{m}}\rp}\,.
\ea
If we want to twine the elliptic genus by an Abelian symmetry that is generated by an element $g$ acting via %with the following action
\be
g: \Phi_i \rightarrow e^{2 \pi \rmi \alpha_i} \Phi_i\,,\qquad  i=1,2,\ldots,d+2\,,
\ee
then this leads to a shift of the original $z$ coordinate (i.e. the second argument) of the $\theta_1$-functions for each $\Phi_i$ by $\alpha_i$. The resulting twined elliptic genus is therefore given by
\ba
\Z^{(g)}_{CY_d}(\tau,z) &=& \Tr_{RR} \lp g\, (-1)^{F_L} y^{J_0} q^{L_0-\frac{d}{8}} (-1)^{F_R} \bar{q}^{\bar L_0-\frac{d}{8}}\rp\cr
&=& \sum_{k,\ell=0}^{m-1} \frac{e^{-2\pi \rmi \ell z}}{m} \prod_{i=1}^{d+2} \frac{\theta_1\lp\tau , \alpha_i+\frac{w_i}{m} (k+\ell \tau +z)-z\rp}{\theta_1\lp\tau , \alpha_i+\frac{w_i}{m} (k+\ell \tau +z)\rp} \cr
&=& \sum_{k,\ell=0}^{m-1} \frac{y^{-\ell}}{m} \prod_{i=1}^{d+2} \frac{\theta_1\lp q, e^{2 \pi \rmi  \lp \alpha_i +\frac{w_i k}{m}\rp} q^{\frac{w_i \ell}{m}} y^{\frac{w_i}{m}-1}\rp}{\theta_1\lp q,e^{2 \pi \rmi  \lp \alpha_i +\frac{w_i k}{m}\rp} q^{\frac{w_i \ell}{m}} y^{\frac{w_i}{m}}\rp}\,.
\ea
The method of \cite{Benini} can also be used for non-Abelian permutation symmetries. In this case one can perform a coordinate transformation that diagonalizes the permutation matrix that acts on the $\Phi_i$. In this new basis the action is then again Abelian and the phase factors are the eigenvalues of the permutation matrix.

\subsection{Calabi-Yau 5-folds}\label{ssec:CY5}
We have seen above that up to an overall constant all CY 5-folds allow for a character expansion that is essentially the same as in Mathieu moonshine. We would now like to understand the potential meaning of this fact better and we do so by studying explicitly the twined elliptic genus for many CY 5-folds. A list of 5\,757\,727 CY 5-folds that can be described by reflexive polytopes  is given on the website \cite{CYwebsite} under ~{\tt 4 folds -> All files}~ in the files ~{\tt 6dRefWH.xxx-xxx.gz}. However, out of these 5\,757\,727 CY 5-folds only 19\,353 are described by transverse polynomials in weighted projective spaces. This is unfortunately required in order to apply our method for calculating the twined elliptic genus, so that we have 19\,353 CY 5-folds at our disposal.

We can now calculate the twined elliptic genus as a power series in $q$ for these examples to get a better understanding of a potential connection to a sporadic group. Based on the elliptic genus in equation \eqref{eq:CY5ell}
\ba
\Z_{CY_5}(\tau,z) &=& -\frac{\chi_{CY_5}}{48} \Bigg[ 22\lp \ch_{5,0,\frac12}^{\N=2}(\tau,z)+\ch_{5,0,-\frac12}^{\N=2}(\tau,z) \rp -2\lp\ch_{5,0,\frac32}^{\N=2}(\tau,z)+\ch_{5,0,-\frac32}^{\N=2}(\tau,z)\rp \cr
&&\qquad \qquad+\sum_{n=1}^\infty A_n \lp \ch_{5,n,\frac32}^{\N=2}(\tau,z)+\ch_{5,n,-\frac32}^{\N=2}(\tau,z) \rp\Bigg]\,,
\ea
we could hope that the elliptic genus for a genuine CY 5-fold transforms like an overall constant times the appropriate function from Mathieu moonshine. This happens in particular for the product CY$_3 \times $K3, where the elliptic genus is the product of a prefactor $\chi_{CY_3}$ and a function that exhibits Mathieu moonshine.

For genuine CY 5-folds we can indeed find this behavior. For the hypersurface in the weighted projective space $\CP^6_{1, 1, 1, 3, 5, 9, 10}$ we have
\ba
\Z_{CY_5}(\tau,z)&=&3556 \cdot \Bigg[ 22\lp \ch_{5,0,\frac12}^{\N=2}(\tau,z)+\ch_{5,0,-\frac12}^{\N=2}(\tau,z) \rp -2\lp\ch_{5,0,\frac32}^{\N=2}(\tau,z)+\ch_{5,0,-\frac32}^{\N=2}(\tau,z)\rp \cr
&& \qquad\qquad +90 \lp \ch_{5,1,\frac32}^{\N=2}(\tau,z)+\ch_{5,1,-\frac32}^{\N=2}(\tau,z) \rp + \ldots\Bigg]\,,
\ea
corresponding to an Euler number of $\chi=-170\,688$. For the $\mathbb{Z}_2$ symmetry
\be\label{eq:CY5Z2}
\mathbb{Z}_2: \left\{ \begin{array}{c} \Phi_{1} \rightarrow -\Phi_{1}\,,\\ \Phi_{2} \rightarrow -\Phi_{2}\,, \end{array} \right.
\ee
we find the twined elliptic genus
\ba
\Z_{CY_5}^{tw,{\rm 2A}}(\tau,z)&=& 14 \cdot \Bigg[ 2\lp \ch_{5,0,\frac12}^{\N=2}(\tau,z)+\ch_{5,0,-\frac12}^{\N=2}(\tau,z) \rp -2\lp\ch_{5,0,\frac32}^{\N=2}(\tau,z)+\ch_{5,0,-\frac32}^{\N=2}(\tau,z)\rp \cr
&& \qquad\ +6 \lp \ch_{5,1,\frac32}^{\N=2}(\tau,z)+\ch_{5,1,-\frac32}^{\N=2}(\tau,z) \rp + \ldots\Bigg]\,,
\ea
which is a twined constant, 14 instead of 3556, multiplied by the 2A series of M$_{24}$ (see appendix \ref{app:CharTab} for the character tables of M$_{12}$ and M$_{24}$ as well as equations \eqref{eq:M24decom} and \eqref{eq:M12decom} for the decomposition of the first few coefficients into irreducible representations). One can decompose the 3556 in a variety of ways into a sum of dimensions of irreducible representations of M$_{24}$ in such a way that one gets 14 upon twining by the 2A element.

So from the above it seems possible that we have an M$_{24}$ symmetry acting on the elliptic genus of this particular CY 5-fold. Since the elliptic genus is an index, this symmetry could arise at certain points in moduli space or everywhere in moduli space or only via symmetry surfing. This possibility extends to other CY 5-folds, where simple low order twinings agree with the expectation. Here we list a few interesting examples:

Applying the same order two twining as the one given in equation \eqref{eq:CY5Z2} to the Calabi-Yau hypersurface in the weighted projective space $\CP^6_{1, 2, 2, 3, 4, 4, 8}$ results again in the 2A element of M$_{24}$, but now with a half integral prefactor of 69/2. This might be seen as a first concern about the appearance of M$_{24}$ but we could remedy this by restricting ourselves to M$_{12}$ instead.

We can likewise study order four twinings by the $\mathbb{Z}_4$ symmetry that acts as
\be\label{eq:CY5Z4}
\mathbb{Z}_4: \left\{ \begin{array}{l} \Phi_{1} \rightarrow \rmi \, \Phi_{1}\,,\\ \Phi_{2} \rightarrow -\rmi\, \Phi_{2}\,. \end{array} \right.
\ee
Here we find for the Calabi-Yau hypersurface in $\CP^6_{1, 1, 1, 1, 4, 4, 4}$ the 4B series with prefactor 42, which is an important answer. Similarly, one can find more specific examples where particular symmetries for particular Calabi-Yau 5-folds give the expected twining function of Mathieu moonshine up to an overall prefactor.

One can perform a more systematic study of twinings of small order for CY 5-folds that are hypersurfaces in weighted projective spaces. We did this for the Calabi-Yau 5-folds that are listed on the website \cite{CYwebsite} in the following manner:
\begin{enumerate}
\item We took the 19\,353 CY 5-folds that are amenable to our method, i.e.~that can be described by a transverse polynomial in the homogeneous coordinates of the ambient weighted projected space.
\item Then we used a very simple code to construct a single transverse polynomial. Due to the simplicity of this code, it found a transverse polynomial only for 18\,880 CY 5-folds.
\item We checked this single transverse polynomial for a $\mathbb{Z}_2$ symmetry and found that 16\,727 of the manifolds have at least one such symmetry.
\item For these 16\,727 manifolds we calculated the twined elliptic genus for a single $\mathbb{Z}_2$ symmetry to zeroth order in $q$. If this calculation took too long, then we aborted it. This led to a set of 13\,642 twined elliptic genera, which we believe correspond to a representative collection of CY 5-folds that are hypersurfaces in weighted projective spaces of low degree.
\end{enumerate}

We find that the order two twining by the $\mathbb{Z}_2$ symmetry always leads to a function that is a linear combination of the 1A and 2A series of M$_{24}$ with prefactors that are integer or half integer. For example, for the Calabi-Yau hypersurface in the weighted projective space $\CP^6_{1, 1, 1, 1, 1, 1, 3}$ we find, when twining by the symmetry in equation \eqref{eq:CY5Z2}, the following twined elliptic genus
\ba
\Z_{CY_5}^{tw}(\tau,z)&=& \frac92 \cdot \Bigg[ 22\lp \ch_{5,0,\frac12}^{\N=2}(\tau,z)+\ch_{5,0,-\frac12}^{\N=2}(\tau,z) \rp -2\lp\ch_{5,0,\frac32}^{\N=2}(\tau,z)+\ch_{5,0,-\frac32}^{\N=2}(\tau,z)\rp \cr
&& \qquad\ +90 \lp \ch_{5,1,\frac32}^{\N=2}(\tau,z)+\ch_{5,1,-\frac32}^{\N=2}(\tau,z) \rp + \mathcal{O}(q^2)\Bigg]\cr
&+& 43 \cdot \Bigg[ 6\lp \ch_{5,0,\frac12}^{\N=2}(\tau,z)+\ch_{5,0,-\frac12}^{\N=2}(\tau,z) \rp -2\lp\ch_{5,0,\frac32}^{\N=2}(\tau,z)+\ch_{5,0,-\frac32}^{\N=2}(\tau,z)\rp \cr
&& \qquad\ -6 \lp \ch_{5,1,\frac32}^{\N=2}(\tau,z)+\ch_{5,1,-\frac32}^{\N=2}(\tau,z) \rp + \mathcal{O}(q^2)\Bigg]\,.
\ea
It follows from standard CFT arguments \cite{MirandaTwining, Gaberdiel:2010ch} that any elliptic genus twined by a group element $g$ has to be a Jacobi form $\phi_{0,m}^g$ with a potentially non-trivial multiplier under
$\Gamma_0(|g|)=\{ ({\tiny\begin{array}{cc} a&b\\c&d \end{array}} ) \in SL(2,\mathbb{Z}) \text{ with } c=0 \mod |g|\}$, where $|g|$ denotes the order of $g$. This means that $\phi_{0,m}^g$ transforms like
\ba
\phi^g_{0,m} \lp \frac{a \tau+b}{c \tau+d},\frac{z}{c \tau+d}\rp &=& e^{\frac{2 \pi \rmi c d}{|g| h_g}}e^{\frac{2\pi \rmi m c
z^2}{c\tau+d}} \phi^g_{0,m}(\tau,z)\,, \qquad\qquad \qquad \,\, \lp \begin{array}{cc} a&b\\
  c&d \end{array} \rp \in \Gamma_0(|g|)
%SL(2,\mathbb{Z}) \text{ and $c=0$ mod $|g|$}
  \,,\cr
\phi^g_{0,m}(\tau, z+\lambda \tau+\mu)&=&(-1)^{2 m(\lambda +\mu)} e^{-2\pi \rmi m (\lambda^2 \tau+2\lambda z)} \phi^g_{0,m}(\tau,z)\,, \qquad \lambda,\mu \in \mathbb{Z} \,.
\ea
Since $c/|g|$ and $d$ are integers we find that the above transformation can only lead to a non-trivial multiplier, i.e. a non-trivial phase in the first transformation law, if $h_g\neq 1$. A list of the $h_g$ for Mathieu moonshine, which is the relevant case for us, is given for example in table 2 in \cite{GannonModule}. In particular one sees there that for $1A$ and $2A$ we have $h_{1A}=h_{2A}=1$, while for 2B one has $h_{2B}=2$, so that the 2B twined elliptic genus has a non-trivial multiplier.

For low order twinings, the twined elliptic genera of CY
5-folds are generated by very few basis elements. This follows from
the fact that this is the case for Mathieu moonshine and K3
manifolds, as discussed for example in \cite{Cheng:2016org}. The
functions that appear for CY 5-folds are simply the product of
$\phi_{0,\frac32}$ and the functions that appear for K3. This is
implied by Lemma 1.4 in \cite{Gritsenko:1999fk}. We are particularly
interested in order two twinings and would like to know whether they
match the expectation of Mathieu moonshine, i.e.~whether the
expansion coefficients conform to the 2A or 2B expectation. On
general grounds twined elliptic genera of order two transform under
$\Gamma_0(2)$ potentially with a non-trivial multiplier. If there is
no non-trivial multiplier, then the answer has to be a linear
combination of the 1A element and the 2A element, since these are the
generators of all modular functions of $\Gamma_0(2)$ without
multiplier and with weight zero and index $m=\frac52$. If there is a
non-trivial multiplier then we expect the answer to be proportional
to the 2B element. However, in all our studies we never found the 2B
element. Whenever we found a twined elliptic genus that looked to
lowest order in $q$ like 2B with a prefactor, then we found by
calculating the next to leading order that it is in fact a linear
combination of 1A and 2A that agrees with 2B at lowest
order. This was expected since we only consider geometric symmetries that cannot lead to a non-trivial multiplier, since the latter arises from the failure of level matching in the $g$-twisted sector. This failure to level match can only occur for symmetries that act asymmetrically on left- and right-movers, which is not the case for our geometric symmetries.\footnote{We thank the referee of this paper for pointing this out to us.} This means all our twined elliptic genera have to be linear combinations of the 1A and 2A elements of $M_{24}$. Since each coefficient in the expansion
of the elliptic genus counts states with a given mass and charge, all
the expansion coefficients of the basis functions have to have
integer coefficients. This means that for our geometric order two
twinings, we expect the answer to be a linear combination with
integer coefficients of the following two basis functions
\ba\label{eq:basis}
f_{1a}(\tau,z) &=& 11\lp \ch_{5,0,\frac12}^{\N=2}(\tau,z)+\ch_{5,0,-\frac12}^{\N=2}(\tau,z) \rp -\lp\ch_{5,0,\frac32}^{\N=2}(\tau,z)+\ch_{5,0,-\frac32}^{\N=2}(\tau,z)\rp \cr
&&+45 \lp \ch_{5,1,\frac32}^{\N=2}(\tau,z)+\ch_{5,1,-\frac32}^{\N=2}(\tau,z) \rp + \ldots\,,\\
f_{2a}(\tau,z) &=& 3 \lp \ch_{5,0,\frac12}^{\N=2}(\tau,z)+\ch_{5,0,-\frac12}^{\N=2}(\tau,z) \rp -\lp\ch_{5,0,\frac32}^{\N=2}(\tau,z)+\ch_{5,0,-\frac32}^{\N=2}(\tau,z)\rp \cr
&& -3 \lp \ch_{5,1,\frac32}^{\N=2}(\tau,z)+\ch_{5,1,-\frac32}^{\N=2}(\tau,z) \rp + \ldots\,.
\ea
These two basis function are 1/2 times the expected 1A and 2A series from Mathieu moonshine, respectively. The factor of 1/2 arises because in Mathieu moonshine all coefficients in the character expansion are even integers. 

\begin{figure}
\centering
\begin{subfigure}{.5\textwidth}
  \centering
  \includegraphics[width=.9\linewidth]{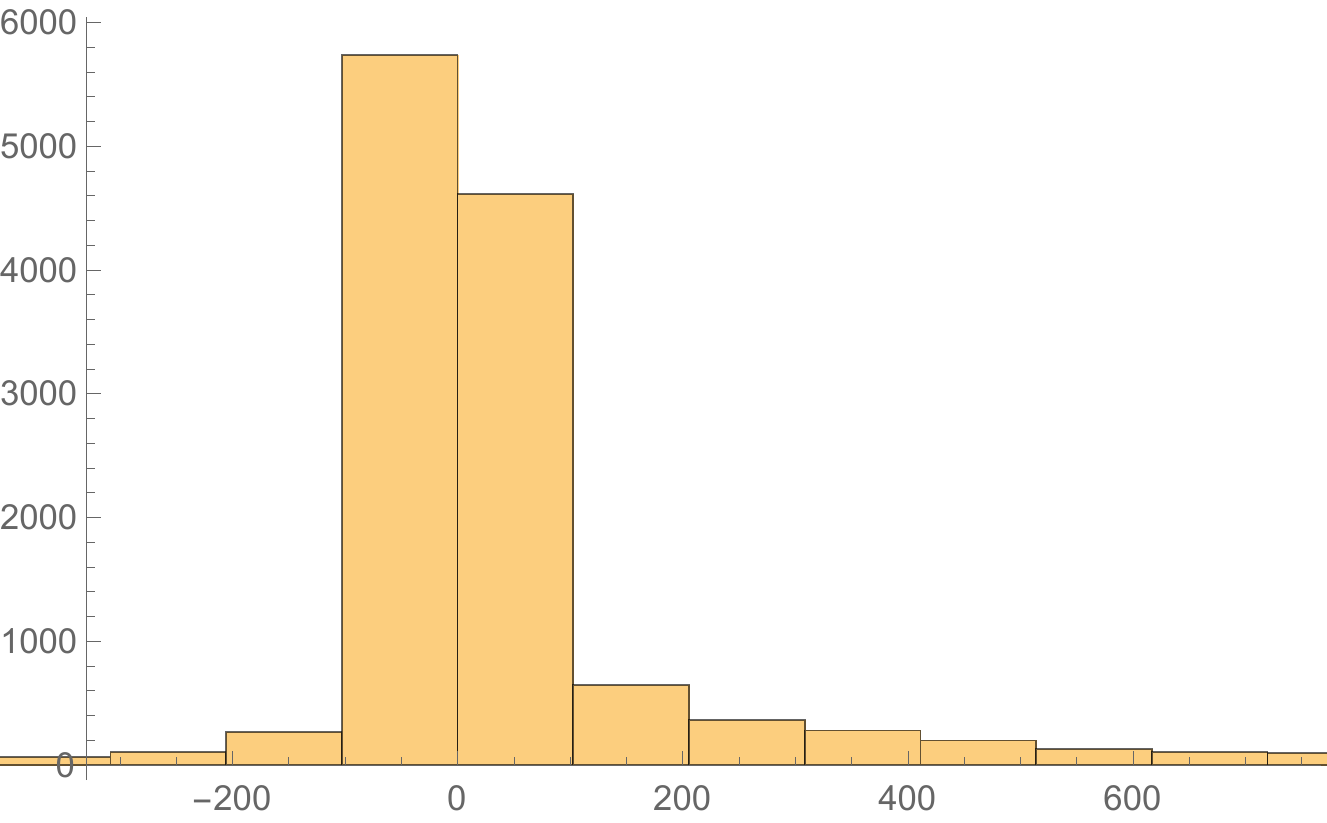}
\end{subfigure}%
\begin{subfigure}{.5\textwidth}
  \centering
  \includegraphics[width=.9\linewidth]{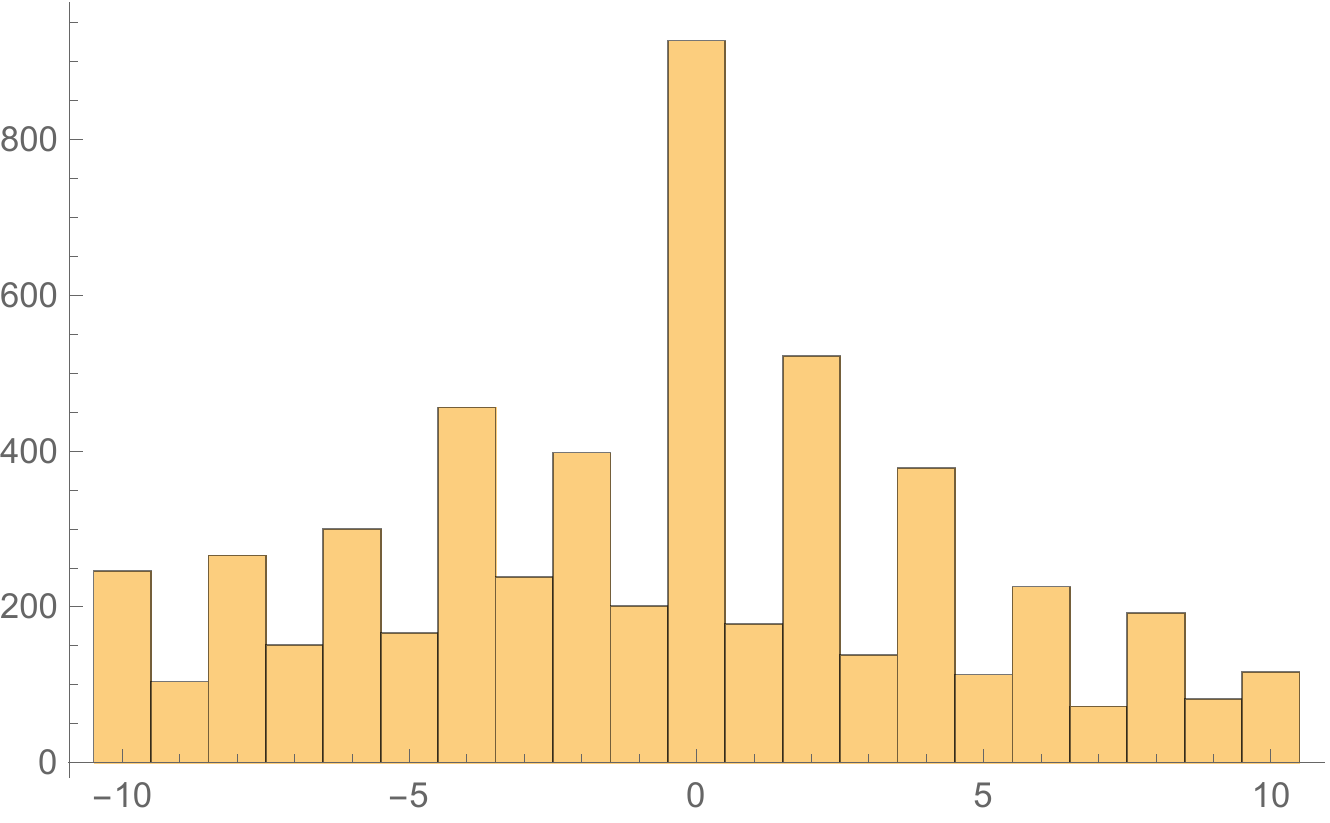}
\end{subfigure}
\caption{A histogram of the coefficients of $f_{1a}$ for the 13\,642 twined elliptic genera. The coefficients peak near zero and as can be seen on the zoomed in histogram on the right this peaking is potentially larger than expected. We also see that even integer coefficients appear substantially more frequent.}
\label{fig:coeff}
\end{figure}

Out of the 13\,642 twined elliptic genera we find in 927 cases an answer that is proportional to $f_{2a}$. For 811 of these the overall coefficient is an even integer and these function are then in agreement with the expectation of an M$_{24}$ symmetry (and for the remaining 116 cases the prefactor is odd and is therefore in agreement with the 2B element in M$_{12}$ which corresponds to the 2A element of M$_{24}$ and therefore to $f_{2a}(\tau,z)$ above). We are now facing the question of whether this is due to pure chance or arises because some of these manifolds actually have a sporadic symmetry. In figure \ref{fig:coeff} we show a histogram of the coefficients of the $f_{1a}$ function for our 13\,642 twined elliptic genera.

We see clearly that the coefficients peak at zero. This peaking near smaller values might be expected when plotting all coefficients as we did on the left side of figure \ref{fig:coeff} since there are a few outliers with rather large values. However, once we zoom in to the region near zero, we see that there is a clear peak at zero. Since a zero coefficient for $f_{1a}$ is roughly twice as likely as a non-zero even integer coefficient it is not quite clear whether we can attritube our 927 instances of a twined elliptic genus proportional to $f_{2a}$ to chance or not. We therefore analyzed these 927 cases further:
\begin{enumerate}
\item We tried to generate a large number of $\mathbb{Z}_2$ symmetries for each of these 927 examples and calculated their elliptic genera for all these different $\mathbb{Z}_2$.
\item For almost all cases with multiple $\mathbb{Z}_2$ symmetries we found the single instance of an elliptic genus that is proportional to 2A and several other order two twined elliptic genera that are linear combinations of $f_{1a}$ and $f_{2a}$. This excludes an action of only M$_{24}$ and supports the explanation that interesting twining genera arose by chance.
\item For cases that had only one obvious $\mathbb{Z}_2$ symmetry we tried higher order twinings by Abelian symmetries. For all cases we found at least one symmetry that leads to a twined elliptic genus that is not consistent with the M$_{24}$ (or M$_{12}$) expectation.
\end{enumerate}

So based on our study of order two (and sometimes also higher order) symmetries we can exclude the possibility that any of the 13\,642 CY 5-folds we studied has a strict M$_{24}$ symmetry. Our results are all consistent with the expectation that these CY 5-folds have small discrete symmetry groups at certain points in moduli space but no large sporadic symmetry. However, we cannot exclude the more exotic possibility of %that these manifolds have
symmetry groups that are even larger than M$_{24}$ and, %in particular based on 
as indicated by our results above, should contain one or several $\mathbb{Z}_2$'s in addition to the M$_{24}$ symmetry group.

Another highly speculative possibility that one could entertain is that generic CY 5-folds have multiple copies of M$_{24}$ as symmetry groups of their superconformal field theories. So instead of interpreting the pre-factor $-\frac{\chi_{CY_5}}{48}$ as a particular sum of irreducible representations of a single M$_{24}$, we could have potentially up to $\left|\frac{\chi_{CY_5}}{48}\right|$ \emph{different} M$_{24}$  symmetries.\footnote{Or analogously up to $\left|\frac{\chi_{CY_5}}{24}\right|$ \emph{different} M$_{12}$ symmetries. We restrict to the M$_{24}$ case here for clarity of the presentation.} A geometric $\mathbb{Z}_2$ symmetry could then correspond to the 1A element in some of the M$_{24}$'s and to the 2A element in other M$_{24}$ groups and the prefactor could be a sum of non-trivial irreducible representations for some of the M$_{24}$ symmetries. This would be consistent with all the results above (except for cases with half integral coefficients that would require us to replace M$_{24}$ by M$_{12}$). While this logical possibility is unmotivated and most likely wrong, it is usually %tricky to really check this
hard to falsify. The only cases where a check is rather straightforward are provided by the most interesting examples of CY 5-folds, namely the ones with Euler numbers $\chi(CY_5)=\pm 24$ and $\chi(CY_5)=\pm 48$. Based on the expansion of their elliptic genus, they could have a single M$_{24}$ symmetry group for $\chi(CY_5)=\pm 48$ or a single M$_{12}$ symmetry group for $\chi(CY_5)=\pm 24$. Unfortunately, scanning through the list of \cite{CYwebsite}, we find that out of the 19\,353 examples only a single one has such a small\footnote{Recall that $\chi(CY_5)$ is always divisible by 24.} Euler number. This is due to the fact that the examples from \cite{CYwebsite} all have a rather small sum of weights $m=\sum_i w_i \leq 200$ and hence mostly a rather large negative Euler number. It is possible to use PALP \cite{Braun:2012vh} to go to larger $m$, but this is rather time consuming and would not lead to a much larger sample of CY 5-folds with small Euler number.

In order to generate a few dozens of examples of CY 5-folds with Euler number $\pm24$ and $\pm48$ we proceeded as follows: First we partitioned an integer $m$ into seven integer weights $w_i$. Then we checked whether the Poincare polynomial
\be
P(x) = \prod_{i=1}^7 \frac{1-x^{m-w_i}}{1-x^{w_i}}
\ee
evaluated at $x=1$ is an integer. This substantially reduces the number of possible examples and is a very fast check. Next we apply the formula 
\be
\chi = \frac{1}{m} \sum_{k=1}^m \sum_{l=1}^m \prod_{gcd(l, k)\cdot \frac{w_i}{m} \in \mathbb{Z}} \frac{w_i-m}{w_i}
\ee
for the Euler number (see \cite{Kreuzer:1992np} for a derivation based on \cite{Vafa:1989ih}) to the remaining cases %all If $\chi$ turns out to be
and proceed only if it results in $\pm24$ or $\pm48$. Then the weights $w_i$ may or may not correspond to a weighted projective space that allows for a CY$_5$ hypersurface that is described by a transverse polynomial. For $7\leq m \leq 600$ we calculated all such weight systems and then checked them explicitly with PALP \cite{Braun:2012vh}, which was easy due to the small number of remaining cases. In this way we were able to generate dozens of examples with small Euler numbers. In order to check whether these examples correspond to different manifolds we calculated all their Hodge numbers. In a few instances we found pairs or triplets of manifolds with the same Hodge numbers, so that the possibility exists that these manifolds are diffeomorphic although they are described by hypersurfaces in different weighted projective spaces. For all of these examples we found all possible geometric $\mathbb{Z}_2$ symmetries using PALP. In table \ref{tab:smallhodge} we list the number of manifolds that we constructed, the number of examples with a $\mathbb{Z}_2$ symmetry and the number of those that can be described by reflexive polytopes.

\begin{table}
\begin{center}
\begin{tabular}{|c|c|c|c|}
  \hline
  Euler number & number of examples & cases with $\mathbb{Z}_2$ symmetries & reflexive cases \\\hline
  $\chi=-48$ &  72 \, (67) & 64\, (59) & 6\,(6) \\\hline
  $\chi=+48$ &  68 \, (59) & 51\, (43)& 4\,(4) \\\hline
  $\chi=-24$ &  32 \, (29) & 26\, (23)& 4\,(4) \\\hline
  $\chi=+24$ & 27 \, (24) & 25\, (22)& 4\,(4) \\\hline
\end{tabular}
\caption{The number of Calabi-Yau 5-folds that we constructed as hypersurfaces in weighted projective spaces. The values in parenthesis give the number of Calabi-Yau manifolds with different Hodge numbers.}\label{tab:smallhodge}
\end{center}
\end{table}

We twined all of the above examples by all of their geometric $\mathbb{Z}_2$ symmetries and found a somewhat surprising result: For the cases of $\chi = + 48$ and $\chi=-48$ we find twined elliptic genera that are proportional to the 2A expectation of M$_{24}$ in 3.4\% and 5.2\% of the cases, respectively. Looking at the histogram plot of the coefficients of $f_{1A}$ we find that zero is not more likely than other small values and there is no peaking as in the right plot in figure \ref{fig:coeff}. Likewise the preference for even coefficients is not apparent. %The surprising feature is
We note, however, the peculiar fact that the cases that give something proportional to 2A have prefactors with absolute values larger than one. Recall that these manifolds have elliptic genera of
\ba
\Z^{\chi=\pm48}_{CY_5}(\tau,z) &=&\mp\Big[ 22\lp \ch_{5,0,\frac12}^{\N=2}(\tau,z)+\ch_{5,0,-\frac12}^{\N=2}(\tau,z) \rp -2\lp\ch_{5,0,\frac32}^{\N=2}(\tau,z)+\ch_{5,0,-\frac32}^{\N=2}(\tau,z)\rp \cr
&&+\sum_{n=1}^\infty A_n \lp \ch_{5,n,\frac32}^{\N=2}(\tau,z)+\ch_{5,n,-\frac32}^{\N=2}(\tau,z) \rp\Big]\cr
&=&\mp 2 f_{1a}(\tau,z)\,,\label{eq:chi48}
\ea
i.e.~a function whose coefficients are all sums of irreducible representations of M$_{24}$ without an overall prefactor. One would therefore have hoped that the twined functions, if they are 2A of M$_{24}$, have no additional prefactor either, i.e~we would have expected to find $\pm2f_{2a}$,  but we find instead \{$-42f_{2a},-38f_{2a},-22f_{2a},-6f_{2a},50f_{2a}$\}, i.e.~we never got the 2A series of M$_{24}$ on the nose. This might also be surprising for another reason: The states counted with sign by the elliptic genus (\ref{eq:ellgenus}) can arise from a cancellation %since such a cancellation would require us to be
only if we are at a special point in moduli space where we have states with the same mass and charge but different statistics. In the absence of such cancellations the twined elliptic genus should never indicate a larger number of states with a given mass and charge than the untwined one (since in the twined case we just count the states that are invariant under the twining). However this is exactly what we observe in some cases. For example for the Calabi-Yau 5-fold that is a hypersurface in the weighted projective space $\mathbb{CP}_{16,17,17,34,58,62,102}^6$ we find the elliptic genus given in equation \eqref{eq:chi48} with a plus sign since this space has $\chi=-48$. Upon twining by a $\mathbb{Z}_2$ symmetry that acts as
\be
\mathbb{Z}_2: \left\{ \begin{array}{c} \Phi_{2} \rightarrow -\Phi_{2}\,,\\ \Phi_{6} \rightarrow -\Phi_{6}\,, \end{array} \right.
\ee
we find the following twined elliptic genus
\ba
\Z^{tw}_{CY_5}(\tau,z) &=&  150\lp \ch_{5,0,\frac12}^{\N=2}(\tau,z)+\ch_{5,0,-\frac12}^{\N=2}(\tau,z) \rp -50\lp\ch_{5,0,\frac32}^{\N=2}(\tau,z)+\ch_{5,0,-\frac32}^{\N=2}(\tau,z)\rp \cr
&&-150 \lp \ch_{5,1,\frac32}^{\N=2}(\tau,z)+\ch_{5,1,-\frac32}^{\N=2}(\tau,z) \rp +\mathcal{O}(q^2) \cr
&=&  50 f_{2a}(\tau,z) \,.\label{eq:48tw}
\ea
For example the 22 states that multiply $\ch_{5,0,\frac12}^{\N=2}(\tau,z)+\ch_{5,0,-\frac12}^{\N=2}(\tau,z)$ in equation \eqref{eq:chi48} have now become 150 states. The 22 untwined states must have arisen from a cancellation between $n+22$ bosonic states and $n$ fermionic states, according to the signs in equation~(\ref{eq:ellgenus}). Upon twining we have $m+150$ bosonic states and $m$ fermionic states. This means that we have $n+22>m+150$ since in the twined case we do not count all states but just the ones invariant under the twining. By comparing equations \eqref{eq:chi48} and \eqref{eq:48tw} we see that in this example there must have been such a cancellation of states at each level since the twined elliptic genus has larger expansion coefficients. This is something that we noticed quite generally and not just for the cases with small Euler number. This can probably  be understood as follows: Generic CY 5-folds have several Hodge numbers that are large compared to 48 so in order to get a small Euler number, these large numbers have to cancel rather precisely. For example for the case above we have the non-trivial Hodge numbers 
\be
h^{1,1} = 25,\quad h^{1,2}=0,\quad h^{1,3}=232,\quad h^{1,4}=259,\quad h^{2,2}=1692,\quad  h^{2,3}=1946.\,
\ee
For any Calabi-Yau 5-fold one has (see page 6 in \cite{Gritsenko:1999fk})
\be
\chi_0=\chi_5=0,\quad \chi_1=\chi_4=-\frac{1}{24}\chi_{CY_5},\quad \chi_2=\chi_3=\frac{11}{24}\chi_{CY_5}\,,
\ee
which is consistent with the Hodge numbers above and $\chi_{CY_5}=-48$. When we twine and thereby remove some states, then this cancellation between large numbers is generically not as good anymore and the expansion coefficients for the twined cases are therefore larger. 

We would also like to note that the ambient weighted projective spaces with six complex dimensions are generically singular and the Calabi-Yau 5-manifolds that are hypersurfaces in these spaces are likewise singular. So our large class of Calabi-Yau hypersurfaces in weighted projective spaces consists of CY 5-folds that are generically singular and that are somewhat special in this sense. However, this by itself does not explain why we do not see a connection to the Mathieu group for any of these cases. For example $T^4/\mathbb{Z}_2$ is a singular limit of K3 and it nevertheless gives rise to many twined series that one expects from Mathieu moonshine. 

Lastly, we discuss the twining by order two elements for the Calabi-Yau manifolds with $\chi=\pm24$ that we constructed. For all of these examples we calculated between one and fourteen $\mathbb{Z}_2$ twinings each but never found a twined elliptic genus that is proportional to $f_{2a}$. So all our answers are linear combinations of $f_{1a}$ and $f_{2a}$ with non-zero coefficients for both functions. It might be noteworthy here that for the cases with $\chi=\pm 48$ we sometimes found order two twined series that are proportional to $f_{1a}$ with a coefficient whose absolute value is larger than 2. Something like this did not occur for the cases with $\chi=\pm24$. We can thus exclude the existence of a strict M$_{24}$ or M$_{12}$ symmetry for the over one hundred examples with $\chi=\pm48$ and $\chi=\pm24$ that we constructed.

\subsection{Calabi-Yau 6-folds}
Above we have seen an interesting connection between the product of two CY 3-folds and a particular instance of umbral moonshine. It is, however, unclear how to get further mileage out of this: The twined elliptic genus of a CY 3-fold is proportional to the untwined elliptic genus since it has essentially a trivial expansion and only the prefactor can change. Let us look, for example, at the quintic, i.e.~the CY 3-fold given by a hypersurface in $\CP_{1,1,1,1,1}^4$. Its elliptic genus is given by %Then we find
\be
\Z_{\mathrm{quintic}}(\tau,z) = -100 \phi_{0,\frac32} =-100 \lp \ch_{3,0,\frac12}^{\N=2}(\tau,z)+ \ch_{3,0,-\frac12}^{\N=2}(\tau,z) \rp\,.
\ee
If we now twine by an order 5 symmetry that acts on the first two coordinates as
\be
g:\quad\Phi_1 \rightarrow e^{2\pi \rmi/5} \Phi_1\,, \quad \Phi_2 \rightarrow e^{-2\pi \rmi/5} \Phi_2\,,
\ee
we find
\be
\Z^{(g)}_{\mathrm{quintic}}(\tau,z) = -5 \phi_{0,\frac32} = -5 \lp \ch_{3,0,\frac12}^{\N=2}(\tau,z)+ \ch_{3,0,-\frac12}^{\N=2}(\tau,z) \rp\,.
\ee
So, as expected, the expansion coefficients that are zero to begin with remain zero and only the one single non-vanishing coefficient can change. This means that a twined CY 3-fold will always have an elliptic genus that is proportional to $\phi_{0,\frac32}$. Hence if we twine the product of two CY 3-folds then we always get the same function, $\phi_{0,\frac32}^2$, just with another prefactor. Therefore from this construction we will never get the expected twined series for elements in $2.AGL_3(2)$. Furthermore, the conformal field theory with target space being the product of two CY 3-folds has $\N=(2,2)$ worldsheet supersymmetry, while the expansion in umbral moonshine is in terms of $\N=4$ characters (cf. equation \eqref{eq:umbral6}).

\section{\label{sec:Gep} A toroidal orbifold and two Gepner models}
Above we have twined a large number of CY 5-folds and found no evidence for a connection between these particular manifolds and the sporadic group M$_{24}$ whose irreducible representations appear in the expansion of the elliptic genus. These results suggest that generic CY 5-folds are not involved in Mathieu moonshine. However, it is still possible that one or several special CY 5-folds have M$_{24}$ as their symmetry group or that they are connected to and could potentially explain Mathieu moonshine in some other way. In the absence of a complete list of CY 5-folds (it is not even known whether there are finitely many) we cannot exclude this possibility. However, we can check particular special cases that have large symmetry groups, such as toroidal orbifolds or Gepner models. For such models it is not only possible to compute the elliptic genus but one can also calculate the Hodge-elliptic genus \cite{Kachru:2016igs, Wendland:2017eiw} or the full partition function. If these models have an actual M$_{24}$ symmetry, then coefficients in the elliptic genus, the Hodge-elliptic genus and the full partition function should be sums of irreducible representations of M$_{24}$.\footnote{We thank Shamit Kachru for mentioning this possibility.} However, in the examples below the coefficients are rather large so guessing or seeing a potential decomposition into irreducible representation of M$_{24}$ is non-trivial. We therefore calculate the twined elliptic genus only.

Toroidal orbifolds $T^{10}/G$ for a discrete group $G$ are abundant and we do not attempt to classify all possible $G$ and calculate corresponding twined elliptic genera. Let us however make a few general comments: Monstrous moonshine and Conway moonshine (and in some way also Mathieu moonshine) are connected to $\mathbb{Z}_2$ orbifolds that change the sign of all coordinates of a torus. However in all these cases the torus involved is $T^n$ with $n$ divisible by 4. This is not the case for us. The $T^{10}/\mathbb{Z}_2$ orbifold with the the sign-changing $\mathbb{Z}_2$ has no holomorphic 5-form and its elliptic genus cannot be expanded in $\N=2$ characters. If the group $G=G_1 \times G_2$ factorizes such that $T^{10}/G = T^4/G_1 \times T^6/G_2$ then we have for $G_1 =\mathbb{Z}_m$, $m=2,3,4,6$ the product of a singular limit of K3 with a complex three dimensional space. In this case we expect trivially the same connection to M$_{24}$ as in the K3 case. So let us look at a different highly symmetric case of a toroidal orbifold of $T^{10}$.
 
We take $G=\mathbb{Z}_2^4$ where the $\mathbb{Z}_2$ are generated by $g_a$, $a=1,2,3,4$ with the following actions on the five complex coordinates of $T^{10}$
\ba
g_1&:& \{z^1,z^2,z^3,z^4,z^5\} \rightarrow \{-z^1,-z^2,z^3,z^4,z^5\}\,,\cr
g_2&:& \{z^1,z^2,z^3,z^4,z^5\} \rightarrow \{z^1,-z^2,-z^3,z^4,z^5\}\,,\cr
g_3&:& \{z^1,z^2,z^3,z^4,z^5\} \rightarrow \{z^1,z^2,-z^3,-z^4,z^5\}\,,\cr
g_4&:& \{z^1,z^2,z^3,z^4,z^5\} \rightarrow \{z^1,z^2,z^3,-z^4,-z^5\}\,.
\ea
This leads to the following elliptic genus
\ba
\Z_{T^{10}/\mathbb{Z}_2^4}(\tau,z)&=&160 \phi_{0,\frac32} \phi_{0,1}\cr
&=& 1280 \bigg[ \frac{\theta_2(\tau,z)^3\theta_3(\tau,z)\theta_4(\tau,z)}{\theta_2(\tau,0)^3\theta_3(\tau,0)\theta_4(\tau,0)}  +\frac{\theta_2(\tau,z)\theta_3(\tau,z)^3\theta_4(\tau,z)}{\theta_2(\tau,0)\theta_3(\tau,0)^3\theta_4(\tau,0)}  \cr
&&\qquad\quad+\frac{\theta_2(\tau,z)\theta_3(\tau,z)\theta_4(\tau,z)^3}{\theta_2(\tau,0)\theta_3(\tau,0)\theta_4(\tau,0)^3}  \bigg]\,.
\ea
We can now twine the above, for example, by the symmetry $g$ that acts as
\be
g: \{z^1,z^2,z^3,z^4,z^5\} \rightarrow \{\rmi z^1,-\rmi z^2,z^3,z^4,z^5\}\,.
\ee
This is an order two symmetry since $g^2=g_1$. The resulting twined elliptic genus is
\ba
\Z^{tw}_{T^{10}/\mathbb{Z}_2^4}(\tau,z)&=& 16 \bigg[ 6 \sum_{j=1}^4 t_j(\tfrac{1}{4}) t_j(-\tfrac{1}{4})t_2(0)t_3(0)t_4(0)\cr
&&\qquad+\ls t_2(0)^2+t_3(0)^2+t_4(0)^2\rs \sum_{k=2}^4 \ls t_1(\tfrac{1}{4})t_k(-\tfrac{1}{4}) +t_k(\tfrac{1}{4})t_1(-\tfrac{1}{4})\rs t_k(0)\cr
&&\qquad+\ls t_2(0)^2+t_3(0)^2+t_4(0)^2\rs \sum_{\substack{k_1\neq k_2\neq k_3\neq k_1\\ k_1,k_2,k_3\in\{2,3,4\}}} t_{k_1}(\tfrac{1}{4}) t_{k_2}(-\tfrac{1}{4}) t_{k_3}(0) \cr
&&\qquad+ \sum_{\substack{k_1\neq k_2\neq k_3\neq k_1\\ k_1,k_2,k_3\in\{2,3,4\}}} \ls t_1(\tfrac{1}{4})t_{k_3}(-\tfrac{1}{4}) +t_{k_3}(\tfrac{1}{4})t_1(-\tfrac{1}{4})+2 t_{k_1}(\tfrac{1}{4})t_{k_2}(-\tfrac{1}{4})\rs\cr
&&\qquad\qquad \qquad\qquad\quad\cdot \lp t_{k_1}(0)^2+t_{k_2}(0)^2\rp t_{k_3}(0)\bigg] \cr
&=& 56 f_{1a} +48 f_{2a}\,,
\ea
where we used the shorthand notation $t_j(x)=\frac{\theta_j(\tau,z+x)}{\theta_j(\tau,x)}$. We see that we again find a linear combination of the $f_{1a}$ and $f_{2a}$ functions, which was required by modular invariance. Note that this particular twining for the K3 orbifold $T^4/\mathbb{Z}_2$ corresponds to the $2A$ element of M$_{24}$ while here there is no connection to M$_{24}$.

Next we would like to study two Gepner models that are highly symmetric, namely the models $(1)^{15}$ and $(2)^{10}$, i.e. we consider the orbifold by $\mathbb{Z}_{k+2}$ of the tensor product of 15 or 10 minimal modes of $A_{k+1}$ type for $k=1$ and $k=2$, respectively. It was explained how to calculate the elliptic genus for the orbifolded tensor product of minimal models in \cite{EdLG, Kawai:1993jk} (this was applied to moonshine in \cite{Cheng:2015rby}). Each $A_{k+1}$ minimal model is obtained from a chiral multiplet $\Phi$ with superpotential $W =\frac{\Phi^{k+2}}{k+2}$ and gives a multiplicative contribution of 
\be
Z_k(\tau,z) = \frac{\theta_1\lp \tau,\tfrac{k+1}{k+2}z\rp}{\theta_1 \lp \tau,\frac{1}{k+2}z\rp}
\ee
to the elliptic genus that, due to the $\mathbb{Z}_{k+2}$ orbifold, is then given by
\be
\Z_{\rm Gepner}(\tau,z) = \frac{1}{k+2}\sum_{a,b=0}^{k+1} e^{\frac{\pi \rmi c}{6} (a+b)} e^{\frac{2 \pi \rmi c}{6} (a^2 \tau+2az)} (Z_k(\tau,z+a \tau+b))^N\,.
\ee
Here $c$ denotes the central charge of the model, which is $c=15$ in our case, and $N$ is the number of minimal models %we tensor together so we have
($N=15$ or $N=10$). From the above we find the elliptic genera
\ba
\Z_{(1)^{15}}(\tau,z) &=& -455 \phi_{0,\frac32}\phi_{0,1}\,,\\
\Z_{(2)^{10}}(\tau,z) &=&-615 \phi_{0,\frac32}\phi_{0,1}\,.
\ea
We can again twine these models by an Abelian symmetry. If the Abelian symmetry multiplies the chiral multiplet by a phase $\Phi \rightarrow e^{2 \pi \rmi \alpha} \Phi$, then the contribution to the twined elliptic genus is
\be
Z_{k,\alpha}(\tau,z) = \frac{\theta_1\lp \tau,\tfrac{k+1}{k+2}z-\alpha\rp }{\theta_1 \lp \tau,\frac{1}{k+2}z+\alpha\rp}\,.
\ee
This leads to the formula
\be
\Z^{tw}_{\rm Gepner}(\tau,z) = \frac{1}{k+2}\sum_{a,b=0}^{k+1} e^{\frac{\pi \rmi c}{6} (a+b)} e^{\frac{2 \pi \rmi c}{6} (a^2 \tau+2az)} \prod_{i=1}^N Z_{k,\alpha_i}(\tau,z+a \tau+b)\,.
\ee
Let us consider the simple $\mathbb{Z}_2$ symmetry that acts by multiplication with a minus sign on the first two chiral multiplets. This results in the %leads to the following two
twined elliptic genera
\ba
\Z^{tw}_{(1)^{15}}(\tau,z) &=& 77 f_{1a} + 110 f_{2a}\,,\\
\Z^{tw}_{(2)^{10}}(\tau,z) &=& 35 f_{1a} + 100 f_{2a}\,.
\ea
So we again find linear combinations of $f_{1a}$ and $f_{2a}$. Note that this particular twining for the Gepner models $(1)^6$ and $(2)^4$ that describe K3 manifolds corresponds in both cases to the $2A$ element of M$_{24}$. So based on this we can conclude that the elliptic genus of K3 seems to certainly exhibit a connection to M$_{24}$ while corresponding models with $c=15$ generically do not share this features.

\section{\label{sec:conclusion} Conclusions}
The observation by Eguchi, Ooguri and Tachikawa that the elliptic genus of the K3 manifold is related to the largest Mathieu group M$_{24}$ \cite{EOT} has started a new era in moonshine in 2010. Since this Mathieu moonshine involves superconformal field theories and Calabi-Yau manifolds of low dimension, it has generated a lot of activity and led to tremendous progress: Mathieu moonshine has been generalized to umbral moonshine and many new related and unrelated moonshine phenomena have been discovered in the last several years. Despite all this progress, however, we still lack a satisfactory explanation of Mathieu moonshine.

Mathieu moonshine connects the Jacobi form $2\phi_{0,1}$ to the Mathieu group M$_{24}$ but up to date it is not settled with absolute certainty that Mathieu moonshine is also connected to the K3 manifold whose elliptic genus is also $2\phi_{0,1}$. The situation should be  compared to Monstrous moonshine, where the $J(\tau)$ function is connected to the Monster group. This connection can be understood via a compactification of the chiral bosonic string theory on the orbifold $\mathbb{R}^{24}/\Lambda_{\rm Leech}/\mathbb{Z}_2$. This orbifold space has the Monster group as its symmetry and the compactification has the $J(\tau)$ function as its partition function. However, since $J(\tau)$ is the unique function (up to an additive constant) that is modular invariant and has a single pole at $\tau=\rmi \infty$, it appears in many other places in physics and mathematics. It seems likely that most or all of these other occurrences have no connection to the Monster group. 

To settle the question of whether K3 is actually related to M$_{24}$ or whether the appearance of $2\phi_{0,1}$ is just an unrelated coincidence, it seems that one could just check the symmetry groups of K3 at different points in moduli space. While these symmetry groups \cite{Gaberdiel} are never exactly M$_{24}$, this does not necessarily mean that there is no connection to Mathieu moonshine. In particular, K3 has a moduli space and different points in moduli space have different symmetry groups. So one should ask which group is relevant for the elliptic genus which counts a subset of states independently of where we are in moduli space. These questions are not easily answered but it has been shown that twined elliptic genera for K3 manifolds give all the McKay-Thompson series of Mathieu moonshine plus several extra twined elliptic genera that one would not have expected based on Mathieu moonshine alone \cite{Cheng:2016org}. At the same time the idea of symmetry surfing has been pursued, in which one tries to find an explicit M$_{24}$ symmetry by combining the symmetry groups at different points in K3 moduli space \cite{Taormina:2011rr, Taormina:2013jza, Taormina:2013mda, Gaberdiel:2016iyz}. In this paper we followed a different path and studied the elliptic genus for other Calabi-Yau manifolds with an eye for potential connections to sporadic groups. While this was not our original goal, we believe that our results further strengthen the connection between the K3 manifold and M$_{24}$ as we explain below.

The elliptic genus for Calabi-Yau manifolds with $d$ complex dimensions is a Jacobi form with weight zero and index $d/2$. These Jacobi forms are generated by very few basis elements and the coefficients can be fixed in terms of very few topological data of the corresponding Calabi-Yau manifold. For CY 3-folds the elliptic genus expanded in terms of $\N=2$ superconformal characters is trivial and has only two non-zero coefficients that are both fixed by the Euler number of the CY 3-fold. For CY$_4$ manifolds the elliptic genus has two basis elements and there seem to exist many interesting connections to sporadic groups that will be explored in a companion paper \cite{CY4}. 

In this paper we mostly focused on CY 5-folds, whose elliptic genus is fixed entirely by the Euler number. Up to an overall prefactor that is proportional to the Euler number, this elliptic genus has the same expansion in terms of $\N=2$ superconformal characters as the elliptic genus of K3, i.e. all the expansion coefficients are given by sums of dimensions of irreducible representations of M$_{24}$. To some extent this puts CY 5-folds on an equal footing with K3, at least CY 5-folds with Euler number $\pm48$ for which the overall prefactor is $\pm 1$, and the expansion of the elliptic genus in superconformal characters for these CY 5-folds is essentially the same as for the K3 manifold. Given this basic observation we set out to investigate what this means by calculating twined elliptic genera for a large number of CY 5-folds with arbitrary Euler number that are hypersurfaces in weighted projective spaces. The reason that we also study CY 5-folds with Euler number different from $\pm 48$, whose elliptic genus expansion agrees with the K3 elliptic genus expansion only up to a prefactor, is that the product spaces K3$\times$CY$_3$ have an elliptic genus that likewise agrees with the K3 elliptic genus expansion only up to a prefactor. However, since these product spaces have an explicit K3 factor they inherit any potential connection to Mathieu moonshine from K3. Therefore it is interesting to check whether generic CY 5-folds exhibit a similar connection as well.

While we found for several CY 5-folds twined elliptic genera that are compatible with the expectation from Mathieu moonshine, it seems possible to attribute these occurrences to pure chance. Given our large set of elliptic genera twined by an order two element, we can make this a little bit more precise. By modular invariance we know that the twined elliptic genus in this case has to be a linear combination of the untwined 1A elliptic genus and the expected twined series corresponding to the 2A element of M$_{24}$, i.e. $\Z_{ell}^{tw}=c_{1a} f_{1a} +c_{2a} f_{2a}$ with generically non-vanishing $c_{1a}$ and $c_{2a}$. This is indeed what we found for the majority of the more than thirteen thousand twined elliptic genera that we computed, but in roughly 900 cases only the 2A element of M$_{24}$ occurred, i.e. we found $c_{1a}=0$. While this seems statistically slightly dominant over other values for $c_{1a}$ (cf. the right side of figure \ref{fig:coeff}), this does not imply that these cases are connected to M$_{24}$. We explicitly showed this by calculating other order two (or higher) twinings for these roughly 900 examples and we always also found some other twined series that are not consistent with the expectation from Mathieu moonshine. 

As mentioned above, CY 5-folds with Euler number $\pm 48$ have essentially the same character expansion of the elliptic genus as the K3 manifold so they deserve particular attention. Since we were only aware of a single example of such a manifold, we constructed over a hundred new examples and studied their twining. We again find order two twined elliptic genera that are linear combinations as above, $\Z_{ell}^{tw}=c_{1a} f_{1a} +c_{2a} f_{2a}$, with generically non-vanishing $c_{1a}$ and $c_{2a}$. As before we found a few cases with $c_{1a}=0$, but this occurred even less often than for the CY 5-folds with other Euler numbers that we studied before. So it seems plausible to attribute these occurrences to pure chance rather than a connection to M$_{24}$. Furthermore, in all these cases we started with the same expansion coefficients for the untwined elliptic genus as for K3 but we then got the 2A series with an additional prefactor that was much larger than one. We attributed this initially surprising observation to generic cancellation between states for CY 5-folds with small Euler numbers (see the end of subsection \ref{ssec:CY5} for details). 

In addition to the large number of CY 5-folds that we studied that were all hypersurfaces in weighted projective space, we also investigated the orbifold $T^{10}/\mathbb{Z}_2^4$ and the Gepner models $(1)^{15}$ and $(2)^{10}$. We calculated the elliptic genus for a simple order two twining and also found for these highly symmetric cases a twined elliptic genus that was of the form $\Z_{ell}^{tw}=c_{1a} f_{1a} +c_{2a} f_{2a}$ with non-vanishing $c_{1a}$ and $c_{2a}$. %Since the number
Lacking a complete list of CY 5-folds, we cannot exclude the possibility that some of them may be related to the sporadic group M$_{24}$. Based on our study, however, we can at least say that generically there seems to be no connection.

Moonshine phenomena are special, rare and surprising connections between particular functions and sporadic groups. In several cases they can be explained by explicit constructions involving conformal field theories with specific target spaces. Given the fact that many CY 5-folds have essentially the same character expansion for their elliptic genus as the K3 surface one might have worried (or hoped) that Mathieu moonshine would potentially not just involve K3 but also a huge number of CY 5-folds. Our results show, based on the twined elliptic genera for CY 5-folds, that this does not seem to be the case. In addition to all the arguments that have been provided for a connection between the K3 surface and Mathieu moonshine in the last several years, we find further evidence from the simple fact that the twined elliptic genera for the K3 manifold agree with the expectation from Mathieu moonshine, while they generically do not agree for the CY 5-folds we studied.

\acknowledgments
We would like to thank M.~Cheng, J.~David, X.~Dong, J.~Duncan, S.~Harrison, J.~Harvey, S.~Kachru, A.~Sen and D.~Whalen for illuminating discussions and collaboration on related projects. AC thanks the HRI, Allahabad for hospitality during part of this project. AK thanks the IIT, Madras for hospitality during the final stage of this project. This work is supported by the Austrian Science Fund (FWF): P 28552.
	
\appendix
	
\section{\label{app:conventions} Conventions}
\subsection{Jacobi theta functions}
The Jacobi theta functions $ \displaystyle \theta_i(\tau,z), \ i = 1, \cdots ,4  $ are defined as
\begin{align}
  \theta_1(\tau,z) &= -\rmi \sum_{n + \frac{1}{2} \in \mathbb{Z}} \left( -1 \right)^{ n - \frac{1}{2} } y^n q^{\frac{n^2}{2}} \nonumber \\ & = -\rmi q^{\frac{1}{8}} \left( y^{\frac{1}{2}} - y^{-\frac{1}{2}} \right) \prod_{n= 1}^\infty \left( 1- q^n \right) \left( 1 - yq^n \right) \left( 1 - y^{-1}q^n \right)\,,\\
  \theta_2(\tau,z) &=\sum_{n + \frac{1}{2} \in \mathbb{Z}}  y^n q^{\frac{n^2}{2}} \nonumber \\
  &=   q^{\frac{1}{8}} \left( y^{\frac{1}{2}} + y^{-\frac{1}{2}} \right) \prod_{n= 1}^\infty \left( 1- q^n \right) \left( 1 + yq^n \right) \left( 1 + y^{-1}q^n \right) \,,\\
  \theta_3(\tau,z) &= \sum_{n \in \mathbb{Z}} y^n q^{\frac{n^2}{2}} \nonumber \\ &= \prod_{n= 1}^\infty \left( 1- q^n \right) \left( 1 + yq^{n - \frac{1}{2}} \right) \left( 1 + y^{-1}q^{n - \frac{1}{2}} \right)\,, \\
  \theta_4(\tau,z) &= \sum_{n \in \mathbb{Z}} \left( -1 \right)^{ n } y^n q^{\frac{n^2}{2}} \nonumber \\ &= \prod_{n= 1}^\infty \left( 1- q^n \right) \left( 1 - yq^{n - \frac{1}{2}} \right) \left( 1 - y^{-1}q^{n - \frac{1}{2}} \right)\,,
\end{align}
where we used $q=e^{2\pi \rmi \tau}$ and $y=e^{2\pi \rmi z}$. We also use the notation for the truncated Jacobi theta function  $ \displaystyle \theta_i(\tau) = \theta_i(\tau, z = 0), \ i = 1,2,3,4.$ The truncated Jacobi theta function can be used to define the Dedekind eta function as
\begin{align}
  \eta(\tau) = \left( \frac{1}{2} \theta_2(\tau) \theta_3(\tau) \theta_4(\tau) \right)^{\frac{1}{3}} = q^{\frac{1}{24}} \prod_{n= 1}^\infty \left( 1 - q^n \right)\,.
\end{align}

\subsection{Eisenstein series}
The Eisenstein series of weights 4 and 6 can likewise be expressed in terms of the $\theta_i(\tau)$ functions
\begin{align}
  E_{4}(\tau) &= \frac{1}{2} \sum_{i = 2}^4 \theta_i(\tau)^8\,, \\
  E_{6}(\tau) &= \frac{1}{2} \lp \theta_4(\tau)^8 \left( \theta_2(\tau)^4 + \theta_3(\tau)^4 \right) + \theta_3(\tau)^8 \left( \theta_4(\tau)^4 - \theta_2(\tau)^4 \right) -\theta_2(\tau)^8 \left( \theta_3(\tau)^4 + \theta_4(\tau)^4 \right) \rp.
\end{align}
The  Eisenstein series have the following Fourier decomposition
\begin{align}
E_4(\tau) &= 1 + 240 \sum_{n = 1}^\infty\frac{n^3 q^n}{1- q^n} = 1 + 240 q + 2160 q^2 + \ldots \,,\\
E_6(\tau) &=1 - 504 \sum_{n = 1}^\infty \frac{n^5 q^n}{1- q^n} = 1 - 504 q - 16632 q^2 +\ldots .
\end{align}

\subsection{Jacobi forms}
The Jacobi forms  $ \displaystyle \phi_{k,m}(\tau,z) $  of weight $ k $ and index $ m $ that we use in this paper are defined as follows:
\ba
\phi_{0,1} (\tau, z) &=& 4 \lp \lp\frac{\theta_2(\tau, z)}{\theta_2(\tau, 0)} \rp^2 +\lp\frac{\theta_3(\tau, z)}{\theta_3(\tau, 0)} \rp^2 +\lp\frac{\theta_4(\tau, z)}{\theta_4(\tau, 0)} \rp^2\rp \label{eq:phi01}\\
&=& \frac{1}{y}+10 +y + \mathcal{O}(q)\,, \cr
\phi_{-2,1}(\tau, z) &=& \frac{\theta_1(\tau, z)^2}{\eta(\tau)^6} \\
&=& -\frac{1}{y}+2-y +\mathcal{O}(q)\,,\cr
\phi_{0, \frac{3}{2}}(\tau,z) &=& 2 \frac{\theta_2(\tau, z)}{\theta_2(\tau, 0)} \frac{\theta_3(\tau, z)}{\theta_3(\tau, 0)} \frac{\theta_4(\tau, z)}{\theta_4(\tau, 0)} \label{eq:phi032}\\
&=& \frac{1}{\sqrt{y}}+\sqrt{y} +\mathcal{O}(q)\,.\nn
\ea

\section{\label{app:characters} Superconformal characters}

\subsection{(Extended) $\N=2$ characters}
For the (extended) $\N=2$ superconformal algebra with central charge $c=3d$, let  $|\Omega\rangle$ denote a highest weight state with eigenvalues $h,\ell$ w.r.t.  $L_0$ and $J_0$.
Writing $\mathcal H_{h,\ell}$ for the representation belonging to $|\Omega\rangle$ we define the (graded) $\N=2$ characters in the Ramond sector through
\be
\ch_{d,h- \frac c {24},\ell} ^{\N=2}(\tau,z) = {\rm tr}_{\mathcal H_{h,\ell}} ((-1)^F q^{L_0-\frac c{24}} e^{2\pi \rmi z J_0})\,,
\ee
where $F$ is the fermion number and $q=e^{2\pi \rmi \tau}$. Below we will also use $y=e^{2\pi\rmi z}$.
In the Ramond sector unitarity  requires $h\geq \frac{c}{24}= \frac d 8$.\\
The characters \cite{Odake:1988bh, Odake:1989dm, Odake:1989ev} are given by (using the conventions of \cite{Eguchi:2010xk}\footnote{Note that our definition of $\theta_1(\tau,z)$ differs by a minus sign from the definition used there.}):
\begin{itemize}
  \item Massles (BPS) representations exist for $h=\frac d 8$; $\ell=\frac d2, \frac d 2 -1 , \frac d 2 - 2, \dots , -(\frac d 2 - 1), -\frac d2$. For $\frac d 2 > \ell\geq0$ they are given by
  \ba
  \ch_{d,0,\ell \geq 0}^{\N=2} (\tau,z)=(-1)^{\ell+\frac d 2}  \frac {(-\rmi) \theta_1(\tau,z)}{\eta(\tau)^3} y^{\ell+\frac 12}
  \sum_{n\in \mathbb Z} q^{\frac {d-1} 2 n^2 + (\ell+\frac 12)n} \frac{\left(-y\right)^{(d-1)n}} {1-y q^n},
  \ea
  and for $\ell=\frac d 2$ one has
  \ba
  \ch_{d,0,\frac d 2}^{\N=2} (\tau,z)=(-1)^{d} \frac {(-\rmi) \theta_1(\tau,z)}{\eta(\tau)^3} y^{\frac{d+1} 2} \sum_{n\in \mathbb Z} q^{\frac {d-1} 2 n^2 +\frac {d+1}2 n}
     \frac{(1-q)\left(-y\right)^{(d-1)n}} {(1-y q^n) (1-y q^{n+1})}.\quad
  \ea
   \item Massive (non-BPS) representations exist for $h> \frac d 8$; $\ell=\frac d 2, \frac d 2 - 1, \dots , -(\frac d 2 - 1), - \frac d 2$  and $\ell\neq 0$ for $d=$ even. For $\ell>0$ we have
  \ba
  \ch_{d,h-\frac c{24},\ell>0}^{\N=2} (\tau,z)=(-1)^{\ell+\frac d 2} q^{h-\frac d 8}
 \frac {\rmi \theta_1(\tau,z)}{\eta(\tau)^3} y^{\ell-\frac 12} \sum_{n\in \mathbb Z} q^{\frac {d-1} 2 n^2 + (\ell-\frac 12)n} \left(-y\right)^{(d-1)n}.\quad
 \ea
\end{itemize}
In both cases the characters for $\ell<0$ are given by
 \be
 \ch_{d,h-\frac c {24},\ell<0}^{\N=2}(\tau,z)= \ch_{d,h-c/24,-\ell>0}^{\N=2}(\tau,-z).
 \ee
%At the unitarity bound $h=\frac d 8$, massive characters decompose into massless ones as, for $\ell\geq 0$,
%  \be
%\lim_{h\searrow \frac d 8}  \ch_{d,h-\frac c {24},\ell+1}^{\N=2}(\tau,z)=\ch_{d,0,\ell+1}^{\N=2}(\tau,z)+\ch_{d,0,\ell}^{\N=2}(\tau,z)\,,
%\ee
%and
% \be
%\lim_{h\searrow \frac d 8}  \ch_{d,h-\frac c {24},\frac d 2}^{\N=2}(\tau,z)=\ch_{d,0,\frac d 2}^{\N=2}(\tau,z)+\ch_{d,0,\frac d 2 -1}^{\N=2}(\tau,z)+\ch_{d,0,-\frac d 2 +1}^{\N=2}(\tau,z).
%\ee
The Witten index of a massless representation is given by
\ba
\ch_{d,0,\ell\geq0}^{\N=2}(\tau,z=0)= \left \{\begin{array}{l  l}
                                                                  (-1)^{\ell+\frac d 2}, & \rm{ for }\: 0\leq\ell< \frac d 2 \,,\\
                                                                  1+(-1)^d, & \rm{ for }\: \ell=\frac d2 \: .
                                                                \end{array}
                                                               \right .
\ea

\subsection{$\N=4$ characters}
Analogously to the $\N=2$  case the (graded) characters of the  $\N=4$ superconformal algebra with central charge $c=3d$, and $d$ even, in the Ramond sector are defined as
\be
\ch_{d,h-\frac c {24},\ell} ^{\N=4}(\tau,z) = {\rm tr}_{\mathcal{H}_{h,\ell}} ((-1)^F q^{L_0-\frac c{24}} e^{4 \pi \rmi z T_0^3  }),
\ee
where  $h$ and  $\ell$ are  the eigenvalues of $L_0$  and $T_0^3$ of the highest weight state belonging to the representation $\mathcal H_{h,\ell}$.
As in the $\N=2$ case unitarity requires $h\geq \frac  d 8$.\\
The characters \cite{Eguchi:1987wf} are given by (using conventions from \cite{Eguchi:2009ux})
\begin{itemize}
  \item Massless representation exist for $h=\frac d 8,\: \ell=0,\frac 12 \dots,\frac d 4 $ and are given by
  \be
  \ch_{d,0,\ell} ^{\N=4}(\tau,z)=\frac {\rmi} {\theta_{1}(\tau,2z)} \frac{\theta_{1}(\tau,z)^2}{\eta(\tau)^3}\sum_{\varepsilon=\pm 1}\sum_{m\in \mathbb Z}
\varepsilon \frac {e^{4\pi \rmi \varepsilon ((\frac d 2 +1)m+\ell)(z+\frac 12)}}{(1-y^{-\varepsilon} q^{-m})^2} q^{(\frac d 2+1)m^2+2\ell m}.
\ee
In particular for $\ell=0$ this may be written as
\be
\ch_{d,0,0} ^{\N=4}(\tau,z) = \frac {-\rmi} {\theta_{1}(\tau,2z)} \frac{\theta_{1}(\tau,z)^2}{\eta(\tau)^3} \sum_{m \in\mathbb Z}  q^{\lp\frac d 2 +1\rp m^2} y^{(d+2) m}\frac{1+y q^m}{1- y q^m}.
\ee
\item Massive representation exist for $h> \frac d 8, \:\ell=\frac 12,1,\dots,\frac d 4$  and are given by
  \be
  \ch_{d,h-\frac c {24},\ell}^{\N=4}(\tau,z)=\rmi q^{h-\frac {2 \ell^2}{d+2}-\frac d8 }\frac{\theta_1(\tau,z)^2}{\theta_1(\tau, 2z) \eta(\tau)^3} \left(\vartheta_{\frac d2 +1,2\ell}(\tau,z+\frac 12)-\vartheta _{\frac d 2+1,-2\ell}(\tau,z+\frac 12)\right),
  \ee
  where
  \be
  \vartheta _{P,a}(\tau,z)=\sum_{n\in \mathbb Z} q^{\frac{(2Pn+a)^2}{4P}} y^{2Pn+a}\,.
  \ee
   \end{itemize}
%  At the unitarity bound $h=\frac d 8$ the massive $\N=4$ characters decompose into massless ones according to
%   \ba
%   \lim_{h\searrow \frac d 8} \ch^{\N=4}_{d,h-\frac c{24},l}(z,\tau)= \ch^{\N=4}_{d,0,l}(z,\tau)+ 2\ch^{\N=4}_{d,0,l-\frac 12}(z,\tau)+\ch^{\N=4}_{d,0,l-1}(z,\tau).
%   \ea
With the help of the $\N=4$ characters combinations of  massless $\N=2$ characters which are even in $z$ can be expressed in the following way
\ba
\ch_{d,0,\frac12}^{\N=2}(\tau,z)+\ch_{d,0,-\frac12}^{\N=2}(\tau,z) &=& (-1)^{\frac{d+1} 2}  \phi_{0,\frac 32} (\tau,z) \ch_{d-3,0,0} ^{\N=4}(\tau,z)\,,\\
\ch_{d,0,\frac32}^{\N=2}(\tau,z)+\ch_{d,0,-\frac32}^{\N=2}(\tau,z) &=& (-1)^{\frac{d+1} 2}  \phi_{0,\frac 32} (\tau,z)\lp \ch_{d-3,0,0} ^{\N=4}(\tau,z)+ \ch_{d-3,0,\frac12} ^{\N=4}(\tau,z)\rp.\qquad\quad
\ea
Likewise the even-$z$ combination of the massive $\N=2$ characters can be written as
\ba
\ch_{d,n,l}^{\N=2}(\tau,z)+\ch_{d,n,-l}^{\N=2}(\tau,z)= (-1)^{2l+\frac {d-1}2} \phi_{0,\frac 32} (\tau,z)  \ch_{d-3,n,\frac 1 2(l-\frac 12)} ^{\N=4} (\tau,z)\,.
\ea

\section{\label{app:CharTab} Character table of M$_{12}$ and M$_{24}$}
\begin{table}[h]
\begin{center}
\caption{The character table of $M_{12}$ where we use the notation $e_{11}=\frac 12 (-1+i \sqrt{11})$.}
\smallskip
\begin{tabular} { c| r r r r r r r r r r r r r r r}
{$[g]$}& 1a  &2a & 2b  &3a  &3b  &4a  &4b  &5a  &6a  &6b  &8a  &8b  &10a  &11a  &11b\\
\hline
$[g^2]$   &1a  &1a  &1a & 3a  &3b  &2b  &2b  &5a  &3b  &3a  &4a  &4b   &5a  &11b  &11a\\
$[g^3]$   &1a & 2a  &2b  &1a  &1a  &4a  &4b  &5a  &2a  &2b  & 8a  &8b  &10a  &11a  &11b\\
$[g^5]$   &1a  &2a  &2b  &3a  &3b  &4a  &4b  &1a  &6a  &6b  &8a  &8b   &2a  &11a  &11b\\
$[g^{11}]$ & 1a & 2a  &2b  &3a  &3b  &4a  &4b  &5a  &6a  &6b  &8a  &8b  &10a   &1a   &1a\\
\hline
$\chi_1$  &      1  & 1  & 1  & 1  & 1  & 1  & 1  & 1  & 1  & 1  & 1  & 1  &  1  &  1  &  1\\
$\chi_2 $  &    11  &-1  & 3  & 2  &-1  &-1  & 3  & 1  &-1  & 0 & -1  & 1  & -1  &  0  &  0\\
$\chi_3  $ &    11  &-1  & 3  & 2 & -1  & 3  &-1  & 1  &-1  & 0  & 1  &-1  & -1   & 0  &  0\\
$\chi_4 $   &   16 &  4  & 0  & -2  & 1  & 0  & 0  & 1  & 1  & 0   &0  & 0   &-1  &  $e_{11}$  & $\bar{e}_{11}$\\
$\chi_5$    &   16  & 4   &0 & -2 &  1  & 0  & 0  & 1  & 1  & 0  & 0 &  0  & -1  & $\bar{e}_{11}$  &  $e_{11}$\\
$\chi_6$    &   45 &  5 & -3  & 0  & 3  & 1  & 1  & 0  &-1  & 0  &-1  &-1  &  0  &  1  &  1\\
$\chi_7$    &   54 &  6  & 6  & 0  & 0   &2  & 2  &-1  & 0  & 0  & 0  & 0  &  1  & -1  & -1\\
$\chi_8 $  &    55  &-5  & 7  & 1  & 1  &-1  &-1  & 0  & 1  & 1  &-1  &-1  &  0  &  0  &  0\\
$\chi_9 $  &    55  &-5  &-1  & 1   &1  & 3  &-1  & 0  & 1  &-1  &-1  & 1  &  0  &  0  &  0\\
$\chi_{10}$   &   55 & -5  &-1  & 1  & 1 &  -1  & 3  & 0  & 1  &-1  & 1  &-1  &  0  &  0  &  0\\
$\chi_{11}$  &    66  & 6  & 2  & 3  & 0  &-2  &-2  & 1  & 0  &-1  & 0  & 0  &  1  &  0  &  0\\
$\chi_{12}$   &   99  &-1  & 3  & 0  & 3  &-1  &-1  &-1  &-1  & 0  & 1  & 1  & -1  &  0  &  0\\
$\chi_{13}$   &  120  & 0 & -8  & 3  & 0 &  0 &  0 &  0  & 0  & 1  & 0  & 0  &  0  & -1  & -1\\
$\chi_{14}$    & 144   &4   &0  & 0  &-3  & 0  & 0  &-1  & 1 & 0  & 0  & 0   &-1   & 1   & 1\\
$\chi_{15}$   &  176  &-4  & 0  &-4  &-1  & 0  & 0  & 1  &-1  & 0  & 0  & 0   & 1   & 0  &  0
\end{tabular}
\end{center}
\end{table}
\clearpage

\begin{sidewaystable}[h]
\begin{center}
\caption{The character table of $M_{24}$ with shorthand notation $e_n=\frac 12 (-1+\rmi\sqrt{n})$.}
\smallskip
\begin{footnotesize}
%\begin{tabular}{c|rrrrrrrrrrrrrrrrr}\toprule
\begin{tabular}{c|rrrrrrrrrrrrrrrrrrrrrrrrrr}%\toprule
 [g]&\text{1a} & \text{2a} & \text{2b} & \text{3a} & \text{3b} & \text{4a} & \text{4b} & \text{4c} & \text{5a} & \text{6a} & \text{6b} & \text{7a} & \text{7b} & \text{8a} & \text{10a} & \text{11a} & \text{12a} & \text{12b} & \text{14a} & \text{14b} & \text{15a} & \text{15b} & \text{21a} & \text{21b} & \text{23a} & \text{23b} \\
\hline
 %\text{1a} & \text{2a} & \text{2b} & \text{3a} & \text{3b} & \text{4a} & \text{4b} & \text{4c} & \text{5a} & \text{6a} & \text{6b} & \text{7a} & \text{7b} & \text{8a} & \text{10a} & \text{11a} & \text{12a} & \text{12b} & \text{14a} & \text{14b} & \text{15a} & \text{15b} & \text{21a} & \text{21b} & \text{23a} & \text{23b} & - \\
  $[g^2]$& \text{1a} & \text{1a} & \text{1a} & \text{3a} & \text{3b} & \text{2a} & \text{2a} & \text{2b} & \text{5a} & \text{3a} & \text{3b} & \text{7a} & \text{7b} & \text{4b} & \text{5a} & \text{11a} & \text{6a} & \text{6b} & \text{7a} & \text{7b} & \text{15a} & \text{15b} & \text{21a} & \text{21b} & \text{23a} & \text{23b} \\
  $[g^3]$& \text{1a} & \text{2a} & \text{2b} & \text{1a} & \text{1a} & \text{4a} & \text{4b} & \text{4c} & \text{5a} & \text{2a} & \text{2b} & \text{7b} & \text{7a} & \text{8a} & \text{10a} & \text{11a} & \text{4a} & \text{4c} & \text{14b} & \text{14a} & \text{5a} & \text{5a} & \text{7b} & \text{7a} & \text{23a} & \text{23b} \\
 $[g^5]$& \text{1a} & \text{2a} & \text{2b} & \text{3a} & \text{3b} & \text{4a} & \text{4b} & \text{4c} & \text{1a} & \text{6a} & \text{6b} & \text{7b} & \text{7a} & \text{8a} & \text{2b} & \text{11a} & \text{12a} & \text{12b} & \text{14b} & \text{14a} & \text{3a} & \text{3a} & \text{21b} & \text{21a} & \text{23b} & \text{23a} \\
  $[g^7]$& \text{1a} & \text{2a} & \text{2b} & \text{3a} & \text{3b} & \text{4a} & \text{4b} & \text{4c} & \text{5a} & \text{6a} & \text{6b} & \text{1a} & \text{1a} & \text{8a} & \text{10a} & \text{11a} & \text{12a} & \text{12b} & \text{2a} & \text{2a} & \text{15b} & \text{15a} & \text{3b} & \text{3b} & \text{23b} & \text{23a} \\
  $[g^{11}]$& \text{1a} & \text{2a} & \text{2b} & \text{3a} & \text{3b} & \text{4a} & \text{4b} & \text{4c} & \text{5a} & \text{6a} & \text{6b} & \text{7a} & \text{7b} & \text{8a} & \text{10a} & \text{1a} & \text{12a} & \text{12b} & \text{14a} & \text{14b} & \text{15b} & \text{15a} & \text{21a} & \text{21b} & \text{23b} & \text{23a} \\
  $[g^{23}]$& \text{1a} & \text{2a} & \text{2b} & \text{3a} & \text{3b} & \text{4a} & \text{4b} & \text{4c} & \text{5a} & \text{6a} & \text{6b} & \text{7a} & \text{7b} & \text{8a} & \text{10a} & \text{11a} & \text{12a} & \text{12b} & \text{14a} & \text{14b} & \text{15a} & \text{15b} & \text{21a} & \text{21b} & \text{1a} & \text{1a} \\
\hline
 $\chi_1$ & 1 & 1 & 1 & 1 & 1 & 1 & 1 & 1 & 1 & 1 & 1 & 1 & 1 & 1 & 1 & 1 & 1 & 1 & 1 & 1 & 1 & 1 & 1 & 1 & 1 & 1 \\
 $\chi_2$ & 23 & 7 & -1 & 5 & -1 & -1 & 3 & -1 & 3 & 1 & -1 & 2 & 2 & 1 & -1 & 1 & -1 & -1 & 0 & 0 & 0 & 0 & -1 & -1 & 0 & 0 \\
 $\chi_3$ & 45 & -3 & 5 & 0 & 3 & -3 & 1 & 1 & 0 & 0 & -1 & $e_7$ & $\bar{e}_7$ & -1 & 0 & 1 & 0 & 1 & $-e_7$ & $-\bar{e}_7$ & 0 & 0 & $e_7$ & $\bar{e}_7$ & -1 & -1 \\
 $\chi_4$ & 45 & -3 & 5 & 0 & 3 & -3 & 1 & 1 & 0 & 0 & -1 & $\bar{e}_7$ & $e_7$ & -1 & 0 & 1 & 0 & 1 & $-\bar{e}_7$ & $-e_7$ & 0 & 0 & $\bar{e}_7$ & $e_7$ & -1 & -1 \\
 $\chi_5$ & 231 & 7 & -9 & -3 & 0 & -1 & -1 & 3 & 1 & 1 & 0 & 0 & 0 & -1 & 1 & 0 & -1 & 0 & 0 & 0 & $e_{15}$ & $\bar{e}_{15}$ & 0 & 0 & 1 & 1 \\
 $\chi_6$ & 231 & 7 & -9 & -3 & 0 & -1 & -1 & 3 & 1 & 1 & 0 & 0 & 0 & -1 & 1 & 0 & -1 & 0 & 0 & 0 & $\bar{e}_{15}$ & $e_{15}$ & 0 & 0 & 1 & 1 \\
 $\chi_7$ & 252 & 28 & 12 & 9 & 0 & 4 & 4 & 0 & 2 & 1 & 0 & 0 & 0 & 0 & 2 & -1 & 1 & 0 & 0 & 0 & -1 & -1 & 0 & 0 & -1 & -1 \\
 $\chi_8$ & 253 & 13 & -11 & 10 & 1 & -3 & 1 & 1 & 3 & -2 & 1 & 1 & 1 & -1 & -1 & 0 & 0 & 1 & -1 & -1 & 0 & 0 & 1 & 1 & 0 & 0 \\
 $\chi_9$ & 483 & 35 & 3 & 6 & 0 & 3 & 3 & 3 & -2 & 2 & 0 & 0 & 0 & -1 & -2 & -1 & 0 & 0 & 0 & 0 & 1 & 1 & 0 & 0 & 0 & 0 \\
 $\chi_{10}$ & 770 & -14 & 10 & 5 & -7 & 2 & -2 & -2 & 0 & 1 & 1 & 0 & 0 & 0 & 0 & 0 & -1 & 1 & 0 & 0 & 0 & 0 & 0 & 0 & $e_{23}$ & $\bar{e}_{23}$ \\
 $\chi_{11}$ & 770 & -14 & 10 & 5 & -7 & 2 & -2 & -2 & 0 & 1 & 1 & 0 & 0 & 0 & 0 & 0 & -1 & 1 & 0 & 0 & 0 & 0 & 0 & 0 & $\bar{e}_{23}$ & $e_{23}$ \\
 $\chi_{12}$ & 990 & -18 & -10 & 0 & 3 & 6 & 2 & -2 & 0 & 0 & -1 & $e_7$ & $\bar{e}_7$ & 0 & 0 & 0 & 0 & 1 & $e_7$ & $\bar{e}_7$ & 0 & 0 & $e_7$ & $\bar{e}_7$ & 1 & 1 \\
 $\chi_{13}$ & 990 & -18 & -10 & 0 & 3 & 6 & 2 & -2 & 0 & 0 & -1 & $\bar{e}_7$ & $e_7$ & 0 & 0 & 0 & 0 & 1 & $\bar{e}_7$ & $e_7$ & 0 & 0 & $\bar{e}_7$ & $e_7$ & 1 & 1 \\
 $\chi_{14}$ & 1035 & 27 & 35 & 0 & 6 & 3 & -1 & 3 & 0 & 0 & 2 & -1 & -1 & 1 & 0 & 1 & 0 & 0 & -1 & -1 & 0 & 0 & -1 & -1 & 0 & 0 \\
 $\chi_{15}$ & 1035 & -21 & -5 & 0 & -3 & 3 & 3 & -1 & 0 & 0 & 1 & $2e_7$ & $2\bar{e}_7$ & -1 & 0 & 1 & 0 & -1 & 0 & 0 & 0 & 0 & $-e_7$ & $-\bar{e}_7$ & 0 & 0 \\
 $\chi_{16}$ & 1035 & -21 & -5 & 0 & -3 & 3 & 3 & -1 & 0 & 0 & 1 & $2\bar{e}_7$ & $2e_7$ & -1 & 0 & 1 & 0 & -1 & 0 & 0 & 0 & 0 & $-\bar{e}_7$ & $-e_7$ & 0 & 0 \\
 $\chi_{17}$ & 1265 & 49 & -15 & 5 & 8 & -7 & 1 & -3 & 0 & 1 & 0 & -2 & -2 & 1 & 0 & 0 & -1 & 0 & 0 & 0 & 0 & 0 & 1 & 1 & 0 & 0 \\
 $\chi_{18}$ & 1771 & -21 & 11 & 16 & 7 & 3 & -5 & -1 & 1 & 0 & -1 & 0 & 0 & -1 & 1 & 0 & 0 & -1 & 0 & 0 & 1 & 1 & 0 & 0 & 0 & 0 \\
 $\chi_{19}$ & 2024 & 8 & 24 & -1 & 8 & 8 & 0 & 0 & -1 & -1 & 0 & 1 & 1 & 0 & -1 & 0 & -1 & 0 & 1 & 1 & -1 & -1 & 1 & 1 & 0 & 0 \\
 $\chi_{20}$ & 2277 & 21 & -19 & 0 & 6 & -3 & 1 & -3 & -3 & 0 & 2 & 2 & 2 & -1 & 1 & 0 & 0 & 0 & 0 & 0 & 0 & 0 & -1 & -1 & 0 & 0 \\
 $\chi_{21}$ & 3312 & 48 & 16 & 0 & -6 & 0 & 0 & 0 & -3 & 0 & -2 & 1 & 1 & 0 & 1 & 1 & 0 & 0 & -1 & -1 & 0 & 0 & 1 & 1 & 0 & 0 \\
 $\chi_{22}$ & 3520 & 64 & 0 & 10 & -8 & 0 & 0 & 0 & 0 & -2 & 0 & -1 & -1 & 0 & 0 & 0 & 0 & 0 & 1 & 1 & 0 & 0 & -1 & -1 & 1 & 1 \\
 $\chi_{23}$ & 5313 & 49 & 9 & -15 & 0 & 1 & -3 & -3 & 3 & 1 & 0 & 0 & 0 & -1 & -1 & 0 & 1 & 0 & 0 & 0 & 0 & 0 & 0 & 0 & 0 & 0 \\
 $\chi_{24}$ & 5544 & -56 & 24 & 9 & 0 & -8 & 0 & 0 & -1 & 1 & 0 & 0 & 0 & 0 & -1 & 0 & 1 & 0 & 0 & 0 & -1 & -1 & 0 & 0 & 1 & 1 \\
 $\chi_{25}$ & 5796 & -28 & 36 & -9 & 0 & -4 & 4 & 0 & 1 & -1 & 0 & 0 & 0 & 0 & 1 & -1 & -1 & 0 & 0 & 0 & 1 & 1 & 0 & 0 & 0 & 0 \\
 $\chi_{26}$ & 10395 & -21 & -45 & 0 & 0 & 3 & -1 & 3 & 0 & 0 & 0 & 0 & 0 & 1 & 0 & 0 & 0 & 0 & 0 & 0 & 0 & 0 & 0 & 0 & -1 & -1 \\
\end{tabular}
%\bottomrule
%\end{tabular}
\end{footnotesize}
\end{center}
\end{sidewaystable}
\clearpage	
	%%%%%%%%%%%%%%%%%%%%%%%%%%%%%%%%%%%%%%%%%%%%%%%%%%%%%%%%%%%
	
\bibliographystyle{JHEP}
\bibliography{refs}

\providecommand{\href}[2]{#2}\begingroup\raggedright\begin{thebibliography}{10}

\bibitem{EOT}
T.~Eguchi, H.~Ooguri and Y.~Tachikawa, \emph{{Notes on the K3 Surface and the
  Mathieu group $M_{24}$}}, {\emph{Exper.Math.} {\bfseries 20} (2011) 91--96},
  [\href{https://arxiv.org/abs/1004.0956}{{\ttfamily 1004.0956}}].

\bibitem{MirandaTwining}
M.~C. Cheng, \emph{{K3 Surfaces, N=4 Dyons, and the Mathieu Group M24}},
  {\emph{Commun.Num.Theor.Phys.} {\bfseries 4} (2010) 623--658},
  [\href{https://arxiv.org/abs/1005.5415}{{\ttfamily 1005.5415}}].

\bibitem{Gaberdiel:2010ch}
M.~R. Gaberdiel, S.~Hohenegger and R.~Volpato, \emph{{Mathieu twining
  characters for K3}},
  \href{http://dx.doi.org/10.1007/JHEP09(2010)058}{\emph{JHEP} {\bfseries 1009}
  (2010) 058}, [\href{https://arxiv.org/abs/1006.0221}{{\ttfamily 1006.0221}}].

\bibitem{Gaberdiel:2010ca}
M.~R. Gaberdiel, S.~Hohenegger and R.~Volpato, \emph{{Mathieu Moonshine in the
  elliptic genus of K3}},
  \href{http://dx.doi.org/10.1007/JHEP10(2010)062}{\emph{JHEP} {\bfseries 1010}
  (2010) 062}, [\href{https://arxiv.org/abs/1008.3778}{{\ttfamily 1008.3778}}].

\bibitem{EguchiTwining}
T.~Eguchi and K.~Hikami, \emph{{Note on Twisted Elliptic Genus of K3 Surface}},
  \href{http://dx.doi.org/10.1016/j.physletb.2010.10.017}{\emph{Phys.Lett.}
  {\bfseries B694} (2011) 446--455},
  [\href{https://arxiv.org/abs/1008.4924}{{\ttfamily 1008.4924}}].

\bibitem{GannonModule}
T.~Gannon, \emph{{Much ado about Mathieu}},
  \href{http://dx.doi.org/10.1016/j.aim.2016.06.014}{\emph{Adv. Math.}
  {\bfseries 301} (2016) 322--358},
  [\href{https://arxiv.org/abs/1211.5531}{{\ttfamily 1211.5531}}].

\bibitem{Govindarajan:2009qt}
S.~Govindarajan and K.~Gopala~Krishna, \emph{{BKM Lie superalgebras from dyon
  spectra in Z(N) CHL orbifolds for composite N}},
  \href{http://dx.doi.org/10.1007/JHEP05(2010)014}{\emph{JHEP} {\bfseries 05}
  (2010) 014}, [\href{https://arxiv.org/abs/0907.1410}{{\ttfamily 0907.1410}}].

\bibitem{Govindarajan:2010fu}
S.~Govindarajan, \emph{{BKM Lie superalgebras from counting twisted CHL
  dyons}}, \href{http://dx.doi.org/10.1007/JHEP05(2011)089}{\emph{JHEP}
  {\bfseries 05} (2011) 089},
  [\href{https://arxiv.org/abs/1006.3472}{{\ttfamily 1006.3472}}].

\bibitem{Duncan:2014vfa}
J.~F.~R. Duncan, M.~J. Griffin and K.~Ono, \emph{{Moonshine}},
  \href{https://arxiv.org/abs/1411.6571}{{\ttfamily 1411.6571}}.

\bibitem{Kachru:2016nty}
S.~Kachru, \emph{{Elementary introduction to Moonshine}},  2016,
  \href{https://arxiv.org/abs/1605.00697}{{\ttfamily 1605.00697}},
  \href{http://inspirehep.net/record/1454427/files/arXiv:1605.00697.pdf}{http://inspirehep.net/record/1454427/files/arXiv:1605.00697.pdf}.

\bibitem{Mukai}
S.~Mukai, \emph{{Finite groups of automorphisms of K3 surfaces and the Mathieu
  group}}, \href{http://dx.doi.org/10.1007/BF01394352}{\emph{Invent. Math.}
  {\bfseries 94} (1988) 183--221}.

\bibitem{Kondo}
S.~Kondo, \emph{Niemeier lattices, mathieu groups, and finite groups of
  symplectic automorphisms of $k3$ surfaces},
  \href{http://dx.doi.org/10.1215/S0012-7094-98-09217-1}{\emph{Duke Math. J.}
  {\bfseries 92} (04, 1998) 593--603}.

\bibitem{Taormina:2011rr}
A.~Taormina and K.~Wendland, \emph{{The overarching finite symmetry group of
  Kummer surfaces in the Mathieu group $M_{24}$}},
  \href{http://dx.doi.org/10.1007/JHEP08(2013)125}{\emph{JHEP} {\bfseries 08}
  (2013) 125}, [\href{https://arxiv.org/abs/1107.3834}{{\ttfamily 1107.3834}}].

\bibitem{Taormina:2013jza}
A.~Taormina and K.~Wendland, \emph{{Symmetry-surfing the moduli space of Kummer
  K3s}}, {\emph{Proc. Symp. Pure Math.} {\bfseries 90} (2015) 129--154},
  [\href{https://arxiv.org/abs/1303.2931}{{\ttfamily 1303.2931}}].

\bibitem{Taormina:2013mda}
A.~Taormina and K.~Wendland, \emph{{A twist in the M24 moonshine story}},
  \href{https://arxiv.org/abs/1303.3221}{{\ttfamily 1303.3221}}.

\bibitem{Gaberdiel:2016iyz}
M.~R. Gaberdiel, C.~A. Keller and H.~Paul, \emph{{Mathieu Moonshine and
  Symmetry Surfing}},
  \href{http://dx.doi.org/10.1088/1751-8121/aa915f}{\emph{J. Phys.} {\bfseries
  A50} (2017) 474002}, [\href{https://arxiv.org/abs/1609.09302}{{\ttfamily
  1609.09302}}].

\bibitem{Gaberdiel}
M.~R. Gaberdiel, S.~Hohenegger and R.~Volpato, \emph{{Symmetries of K3 sigma
  models}}, {\emph{Commun.Num.Theor.Phys.} {\bfseries 6} (2012) 1--50},
  [\href{https://arxiv.org/abs/1106.4315}{{\ttfamily 1106.4315}}].

\bibitem{Cheng:2016org}
M.~C.~N. Cheng, S.~M. Harrison, R.~Volpato and M.~Zimet, \emph{{K3 String
  Theory, Lattices and Moonshine}},
  \href{https://arxiv.org/abs/1612.04404}{{\ttfamily 1612.04404}}.

\bibitem{Paquette:2017gmb}
N.~M. Paquette, R.~Volpato and M.~Zimet, \emph{{No More Walls! A Tale of
  Modularity, Symmetry, and Wall Crossing for 1/4 BPS Dyons}},
  \href{http://dx.doi.org/10.1007/JHEP05(2017)047}{\emph{JHEP} {\bfseries 05}
  (2017) 047}, [\href{https://arxiv.org/abs/1702.05095}{{\ttfamily
  1702.05095}}].

\bibitem{Kachru:2016ttg}
S.~Kachru, N.~M. Paquette and R.~Volpato, \emph{{3D String Theory and Umbral
  Moonshine}}, \href{http://dx.doi.org/10.1088/1751-8121/aa6e07}{\emph{J.
  Phys.} {\bfseries A50} (2017) 404003},
  [\href{https://arxiv.org/abs/1603.07330}{{\ttfamily 1603.07330}}].

\bibitem{umbralone}
M.~C.~N. Cheng, J.~F.~R. Duncan and H.~J. A., \emph{{Umbral Moonshine}},
  \href{http://dx.doi.org/10.4310/CNTP.2014.v8.n2.a1}{\emph{Commun. Num. Theor.
  Phys.} {\bfseries 08} (2014) 101--242},
  [\href{https://arxiv.org/abs/1204.2779}{{\ttfamily 1204.2779}}].

\bibitem{umbraltwo}
M.~C.~N. Cheng, J.~F.~R. Duncan and J.~A. Harvey, \emph{{Umbral Moonshine and
  the Niemeier Lattices}},  \href{https://arxiv.org/abs/1307.5793}{{\ttfamily
  1307.5793}}.

\bibitem{Kawai:1993jk}
T.~Kawai, Y.~Yamada and S.-K. Yang, \emph{{Elliptic genera and N=2
  superconformal field theory}},
  \href{http://dx.doi.org/10.1016/0550-3213(94)90428-6}{\emph{Nucl. Phys.}
  {\bfseries B414} (1994) 191--212},
  [\href{https://arxiv.org/abs/hep-th/9306096}{{\ttfamily hep-th/9306096}}].

\bibitem{CY4}
to~appear.

\bibitem{bookZagier}
M.~M. Eichler and D.~Zagier, \emph{The theory of Jacobi forms}.
\newblock Boston Birkh{\"a}user, 1985.

\bibitem{Gritsenko:1999fk}
V.~Gritsenko, \emph{{Elliptic genus of Calabi-Yau manifolds and Jacobi and
  Siegel modular forms}},  \href{https://arxiv.org/abs/math/9906190}{{\ttfamily
  math/9906190}}.

\bibitem{Eguchi:2010xk}
T.~Eguchi and K.~Hikami, \emph{{N=2 Superconformal Algebra and the Entropy of
  Calabi-Yau Manifolds}},
  \href{http://dx.doi.org/10.1007/s11005-010-0387-3}{\emph{Lett. Math. Phys.}
  {\bfseries 92} (2010) 269--297},
  [\href{https://arxiv.org/abs/1003.1555}{{\ttfamily 1003.1555}}].

\bibitem{Edold}
E.~Witten, \emph{{Elliptic Genera and Quantum Field Theory}},
  \href{http://dx.doi.org/10.1007/BF01208956}{\emph{Commun.Math.Phys.}
  {\bfseries 109} (1987) 525}.

\bibitem{EOTY}
T.~Eguchi, H.~Ooguri, A.~Taormina and S.-K. Yang, \emph{{Superconformal
  Algebras and String Compactification on Manifolds with SU(N) Holonomy}},
  \href{http://dx.doi.org/10.1016/0550-3213(89)90454-9}{\emph{Nucl.Phys.}
  {\bfseries B315} (1989) 193}.

\bibitem{Eguchi:2013es}
T.~Eguchi and K.~Hikami, \emph{{Enriques moonshine}},
  \href{http://dx.doi.org/10.1088/1751-8113/46/31/312001}{\emph{J. Phys.}
  {\bfseries A46} (2013) 312001},
  [\href{https://arxiv.org/abs/1301.5043}{{\ttfamily 1301.5043}}].

\bibitem{Cheng:2013kpa}
M.~C. Cheng, X.~Dong, J.~Duncan, J.~Harvey, S.~Kachru and T.~Wrase,
  \emph{{Mathieu Moonshine and N=2 String Compactifications}},
  \href{http://dx.doi.org/10.1007/JHEP09(2013)030}{\emph{JHEP} {\bfseries 1309}
  (2013) 030}, [\href{https://arxiv.org/abs/1306.4981}{{\ttfamily 1306.4981}}].

\bibitem{Wrase:2014fja}
T.~Wrase, \emph{{Mathieu moonshine in four dimensional $\mathcal{N}=1$
  theories}}, \href{http://dx.doi.org/10.1007/JHEP04(2014)069}{\emph{JHEP}
  {\bfseries 04} (2014) 069},
  [\href{https://arxiv.org/abs/1402.2973}{{\ttfamily 1402.2973}}].

\bibitem{Paquette:2014rma}
N.~M. Paquette and T.~Wrase, \emph{{Comments on M$_{24}$ representations and
  $CY_3$ geometries}},
  \href{http://dx.doi.org/10.1007/JHEP11(2014)155}{\emph{JHEP} {\bfseries 11}
  (2014) 155}, [\href{https://arxiv.org/abs/1409.1540}{{\ttfamily 1409.1540}}].

\bibitem{Datta:2015hza}
S.~Datta, J.~R. David and D.~Lust, \emph{{Heterotic string on the CHL orbifold
  of K3}}, \href{http://dx.doi.org/10.1007/JHEP02(2016)056}{\emph{JHEP}
  {\bfseries 02} (2016) 056},
  [\href{https://arxiv.org/abs/1510.05425}{{\ttfamily 1510.05425}}].

\bibitem{Chattopadhyaya:2016xpa}
A.~Chattopadhyaya and J.~R. David, \emph{{${\cal N}=2$ heterotic string
  compactifications on orbifolds of $K3\times T^2$}},
  \href{http://dx.doi.org/10.1007/JHEP01(2017)037}{\emph{JHEP} {\bfseries 01}
  (2017) 037}, [\href{https://arxiv.org/abs/1611.01893}{{\ttfamily
  1611.01893}}].

\bibitem{Chattopadhyaya:2017ews}
A.~Chattopadhyaya and J.~R. David, \emph{{Dyon degeneracies from Mathieu
  moonshine symmetry}},
  \href{http://dx.doi.org/10.1103/PhysRevD.96.086020}{\emph{Phys. Rev.}
  {\bfseries D96} (2017) 086020},
  [\href{https://arxiv.org/abs/1704.00434}{{\ttfamily 1704.00434}}].

\bibitem{Eguchi:2012ye}
T.~Eguchi and K.~Hikami, \emph{{N=2 Moonshine}},
  \href{http://dx.doi.org/10.1016/j.physletb.2012.09.037}{\emph{Phys. Lett.}
  {\bfseries B717} (2012) 266--273},
  [\href{https://arxiv.org/abs/1209.0610}{{\ttfamily 1209.0610}}].

\bibitem{Cheng:2014owa}
M.~C.~N. Cheng, X.~Dong, J.~F.~R. Duncan, S.~Harrison, S.~Kachru and T.~Wrase,
  \emph{{Mock Modular Mathieu Moonshine Modules}},
  \href{https://arxiv.org/abs/1406.5502}{{\ttfamily 1406.5502}}.

\bibitem{Cheng:2015fha}
M.~C.~N. Cheng, S.~M. Harrison, S.~Kachru and D.~Whalen, \emph{{Exceptional
  Algebra and Sporadic Groups at c=12}},
  \href{https://arxiv.org/abs/1503.07219}{{\ttfamily 1503.07219}}.

\bibitem{Frenkel1985}
I.~B. Frenkel, J.~Lepowsky and A.~Meurman, \emph{A Moonshine Module for the
  Monster}, pp.~231--273.
\newblock Springer US, New York, NY, 1985.
\newblock 10.1007/978-1-4613-9550-8\_12.

\bibitem{Duncan}
J.~F. {Duncan}, \emph{{Super-moonshine for Conway's largest sporadic group}},
  {\emph{ArXiv Mathematics e-prints} (Feb., 2005) },
  [\href{https://arxiv.org/abs/math/0502267}{{\ttfamily math/0502267}}].

\bibitem{umbralproof}
J.~F.~R. Duncan, M.~J. Griffin and K.~Ono, \emph{{Proof of the Umbral Moonshine
  Conjecture}},  \href{https://arxiv.org/abs/1503.01472}{{\ttfamily
  1503.01472}}.

\bibitem{Benini}
F.~Benini, R.~Eager, K.~Hori and Y.~Tachikawa, \emph{{Elliptic genera of
  two-dimensional N=2 gauge theories with rank-one gauge groups}},
  \href{http://dx.doi.org/10.1007/s11005-013-0673-y}{\emph{Lett. Math. Phys.}
  {\bfseries 104} (2014) 465--493},
  [\href{https://arxiv.org/abs/1305.0533}{{\ttfamily 1305.0533}}].

\bibitem{Harrison:2013bya}
S.~Harrison, S.~Kachru and N.~M. Paquette, \emph{{Twining Genera of (0,4)
  Supersymmetric Sigma Models on K3}},
  \href{http://dx.doi.org/10.1007/JHEP04(2014)048}{\emph{JHEP} {\bfseries 04}
  (2014) 048}, [\href{https://arxiv.org/abs/1309.0510}{{\ttfamily 1309.0510}}].

\bibitem{CYwebsite}
M.~Kreuzer and H.~Skarke, ``{Calabi-Yau data}.''
  \url{http://hep.itp.tuwien.ac.at/~kreuzer/CY/}.

\bibitem{Braun:2012vh}
A.~P. Braun, J.~Knapp, E.~Scheidegger, H.~Skarke and N.-O. Walliser,
  \emph{{PALP - a User Manual}},  in \emph{Strings, gauge fields, and the
  geometry behind: The legacy of Maximilian Kreuzer} (A.~Rebhan, L.~Katzarkov,
  J.~Knapp, R.~Rashkov and E.~Scheidegger, eds.), pp.~461--550.
\newblock 2012.
\newblock \href{https://arxiv.org/abs/1205.4147}{{\ttfamily 1205.4147}}.
\newblock \href{http://dx.doi.org/10.1142/9789814412551_0024}{DOI}.

\bibitem{Kreuzer:1992np}
M.~Kreuzer and H.~Skarke, \emph{{No mirror symmetry in Landau-Ginzburg
  spectra!}}, \href{http://dx.doi.org/10.1016/0550-3213(92)90547-O}{\emph{Nucl.
  Phys.} {\bfseries B388} (1992) 113--130},
  [\href{https://arxiv.org/abs/hep-th/9205004}{{\ttfamily hep-th/9205004}}].

\bibitem{Vafa:1989ih}
C.~Vafa, \emph{{Quantum Symmetries of String Vacua}},
  \href{http://dx.doi.org/10.1142/S0217732389001842}{\emph{Mod. Phys. Lett.}
  {\bfseries A4} (1989) 1615}.

\bibitem{Kachru:2016igs}
S.~Kachru and A.~Tripathy, \emph{{The Hodge-elliptic genus, spinning BPS
  states, and black holes}},
  \href{http://dx.doi.org/10.1007/s00220-017-2910-1}{\emph{Commun. Math. Phys.}
  {\bfseries 355} (2017) 245--259},
  [\href{https://arxiv.org/abs/1609.02158}{{\ttfamily 1609.02158}}].

\bibitem{Wendland:2017eiw}
K.~Wendland, \emph{{Hodge-elliptic genera and how they govern K3 theories}},
  \href{https://arxiv.org/abs/1705.09904}{{\ttfamily 1705.09904}}.

\bibitem{EdLG}
E.~Witten, \emph{{On the Landau-Ginzburg description of N=2 minimal models}},
  \href{http://dx.doi.org/10.1142/S0217751X9400193X}{\emph{Int.J.Mod.Phys.}
  {\bfseries A9} (1994) 4783--4800},
  [\href{https://arxiv.org/abs/hep-th/9304026}{{\ttfamily hep-th/9304026}}].

\bibitem{Cheng:2015rby}
M.~C.~N. Cheng, F.~Ferrari, S.~M. Harrison and N.~M. Paquette,
  \emph{{Landau-Ginzburg Orbifolds and Symmetries of K3 CFTs}},
  \href{http://dx.doi.org/10.1007/JHEP01(2017)046}{\emph{JHEP} {\bfseries 01}
  (2017) 046}, [\href{https://arxiv.org/abs/1512.04942}{{\ttfamily
  1512.04942}}].

\bibitem{Odake:1988bh}
S.~Odake, \emph{{Extension of $N=2$ Superconformal Algebra and Calabi-Yau
  Compactification}},
  \href{http://dx.doi.org/10.1142/S021773238900068X}{\emph{Mod. Phys. Lett.}
  {\bfseries A4} (1989) 557}.

\bibitem{Odake:1989dm}
S.~Odake, \emph{{Character Formulas of an Extended Superconformal Algebra
  Relevant to String Compactification}},
  \href{http://dx.doi.org/10.1142/S0217751X90000428}{\emph{Int. J. Mod. Phys.}
  {\bfseries A5} (1990) 897}.

\bibitem{Odake:1989ev}
S.~Odake, \emph{{C = 3-$d$ Conformal Algebra With Extended Supersymmetry}},
  \href{http://dx.doi.org/10.1142/S0217732390000640}{\emph{Mod. Phys. Lett.}
  {\bfseries A5} (1990) 561}.

\bibitem{Eguchi:1987wf}
T.~Eguchi and A.~Taormina, \emph{{Character Formulas for the $N=4$
  Superconformal Algebra}},
  \href{http://dx.doi.org/10.1016/0370-2693(88)90778-2}{\emph{Phys. Lett.}
  {\bfseries B200} (1988) 315}.

\bibitem{Eguchi:2009ux}
T.~Eguchi and K.~Hikami, \emph{{N=4 Superconformal Algebra and the Entropy of
  HyperKahler Manifolds}},
  \href{http://dx.doi.org/10.1007/JHEP02(2010)019}{\emph{JHEP} {\bfseries 02}
  (2010) 019}, [\href{https://arxiv.org/abs/0909.0410}{{\ttfamily 0909.0410}}].

\end{thebibliography}\endgroup
	
\end{document}